\documentclass[%
preprint,
nofootinbib,
amsmath,amssymb,
aps
]{revtex4-2}

\usepackage{graphicx}
\usepackage{dcolumn}
\usepackage{bm}
\usepackage{slashed}
\usepackage{bbold}
\usepackage{enumitem}
\usepackage{graphicx}
\usepackage{subcaption}
\usepackage[T1]{fontenc}
\usepackage[utf8]{inputenc}
\usepackage[force]{feynmp-auto}
\usepackage{float}
\usepackage{multirow}
\usepackage{subdepth} 
\usepackage{booktabs} 

\newcommand{\picwidth}{7cm}
\newcommand{\coloredmath}[2]{%
	\newcommand{#1}{\ensuremath{{{#2}}}}%
}%

\coloredmath{\br}{\operatorname{BR}}
\coloredmath{\rate}{\operatorname{CR}}
\coloredmath{\reals}{\mathbb{R}}
\coloredmath{\im}{\operatorname{Im}}
\coloredmath{\sign}{\operatorname{sign}}

\coloredmath{\kplus}{K^+\to\pi^+\nu\bar\nu}
\coloredmath{\dzero}{D^0\to\mu^+\mu^-}
\coloredmath{\am}{\Delta a_\mu}
\coloredmath{\meg}{\mu\to e\gamma}
\coloredmath{\teg}{\tau\to e\gamma}
\coloredmath{\tmg}{\tau\to \mu\gamma}
\coloredmath{\twodecays}{\ell_i\to\ell_j\gamma}
\coloredmath{\threedecays}{\ell_i\to\ell_j\ell_k^{\phantom{c}}\ell_k^c}
\coloredmath{\meee}{\mu\to3e}
\coloredmath{\teee}{\tau\to 3e}
\coloredmath{\tmmm}{\tau\to 3\mu}
\coloredmath{\tmee}{\tau\to \mu ee}
\coloredmath{\temm}{\tau\to e\mu\mu}
\coloredmath{\mec}{\mu Al\to eAl}
\coloredmath{\mecold}{\mu Au\to eAu}

\coloredmath{\deltaqed}{\delta_{\text{QED}}}
\newcommand{\ratio}[1]{\ensuremath{{k_{#1}}}}
\coloredmath{\gev}{\text{GeV}}
\coloredmath{\tev}{\text{TeV}}
\newcommand{\code}[1]{\texttt{#1}}
\coloredmath{\hc}{h.c.}
\coloredmath{\lhs}{\text{lhs.}}
\coloredmath{\rhs}{\text{rhs.}}
\newcommand{\norm}[1]{|#1|^2}

\newcommand{\ms}[1]{\ensuremath{{m_{S_1}^{#1}}}}
\newcommand{\Lcoupling}[1]{\ensuremath{{\lambda^{#1}_{L}}}}
\newcommand{\Rcoupling}[1]{\ensuremath{{\lambda^{#1}_{R}}}}
\newcommand{\LRcouplings}[1]{\ensuremath{{\lambda^{#1}_{L, R}}}}

\begin{document}	
\title{Constraint on scalar leptoquark from low energy leptonic observables}
	
\author{Uladzimir Khasianevich}
\email{uladzimir.khasianevich@tu-dresden.de}
\author{Dominik St\"ockinger}
\email{dominik.stoeckinger@tu-dresden.de}
\author{Hyejung St\"ockinger-Kim}
\email{hyejung.stoeckinger-kim@tu-dresden.de}
\author{Johannes W\"unsche}
\email{johannes.wuensche@tu-dresden.de}
\affiliation{Institut f\"ur Kern- und Teilchenphysik, TU Dresden, Zellescher Weg 19, 01069 Dresden, Germany}
	
\begin{abstract}
We consider the flavor structure
of the  $S_1$ leptoquark model and derive conservative constraints on the 
elements of the left- and right-handed coupling
matrices in a number of representative scenarios. We focus on the cases where the muon $g-2$ deviation is
explained by real muon couplings to either the top-quark or to the charm-quark or
to all up-type quarks. The most significant constraints arise
from charged lepton flavor violating decays of the muon and the $\tau$
lepton and from the $\mu$--$e$ conversion process.
Kaon decays and perturbativity provide further constraints. We find
strong constraints on almost all coupling matrix elements, implying a
very hierarchical matrix structure, where individual entries must
differ by at least 4 orders of magnitude.
The \texttt{FlexibleSUSY} program was used with appropriate	model files incorporating the parameterization of the couplings in the up-type mass diagonal basis. 
The expressions for the leptonic observables were generated and cross-checked with the help of the \texttt{NPointFunctions} extension
of the \texttt{FlexibleSUSY} program.
\end{abstract}
\maketitle
	
\tableofcontents
	
\section{Introduction}\label{sec:intro}
Low-energy lepton precision physics provides an excellent probe of
fundamental interactions with the potential of discovering new
physics beyond the Standard Model (SM) and shedding light on the origin
of mass and flavor. The anomalous magnetic moment of the muon
$a_\mu$ is a flavor- and CP-conserving observable which corresponds
to a chirality-flipping dipole operator. There is a longstanding discrepancy between the experimental determination at the BNL and
Fermilab measurements and the SM theory prediction:%
\footnote{\label{fn:fnalupdate}%
Since the release of this paper there has been an update on the experimental average of the anomalous magnetic moment of the muon
	 from Run-2 of the FNAL experiment $\Delta a_\mu^{\text{2023}}$, see Ref.~\cite{Muong-2:2023cdq}. This paper represents the measurement from before this latest update.
}
\begin{align}
	\Delta a_\mu^{\text{2021}}=a_\mu^{\text{Exp,
			2021}}-a_\mu^{\text{SM}}=(25.1 \pm
	5.9)\cdot10^{-10}\,.
	\label{eq:Damu}
\end{align}
This value is based on the Fermilab Run-1 result \cite{Muong-2:2021ojo}, the
Brookhaven result \cite{Muong-2:2006rrc}, and the Standard Model White Paper \cite{Aoyama:2020ynm},
which itself uses results from original references \cite{Aoyama:2012wk,Aoyama:2019ryr,Czarnecki:2002nt,Gnendiger:2013pva,Davier:2017zfy,Keshavarzi:2018mgv,Colangelo:2018mtw,Hoferichter:2019mqg,Davier:2019can,Keshavarzi:2019abf,Kurz:2014wya,Melnikov:2003xd,Masjuan:2017tvw,Colangelo:2017fiz,Hoferichter:2018kwz,Gerardin:2019vio,Bijnens:2019ghy,Colangelo:2019uex,Pauk:2014rta,Danilkin:2016hnh,Jegerlehner:2017gek,Knecht:2018sci,Eichmann:2019bqf,Roig:2019reh,Blum:2019ugy,Colangelo:2014qya}.
After the White Paper \cite{Aoyama:2020ynm}, several lattice gauge theory results
\cite{Borsanyi:2020mff,Ce:2022kxy,Alexandrou:2022amy,Blum:2023qou} and
the CMD-3 measurement \cite{CMD-3:2023alj} of $e^+e^-\to\text{hadrons}$ are
in tension with earlier results and tend to prefer higher values of
the hadronic vacuum polarization contributions to $a_\mu$. Taking
those results at face value would reduce $\Delta a_\mu^{\text{2021}}$ to about
half its quoted value, but scrutiny of these results is ongoing, and further progress on the hadronic vacuum polarization contributions is expected in the coming years \cite{Colangelo:2022jxc}. Furthermore, more precise experimental determinations of $\am$ based on Run-2/3 data and later on Run-4/5/6 data from the Fermilab experiment are in preparation. In view of this progress it remains relevant to ask which scenarios for physics beyond the SM could explain a deviation as large as Eq.~\eqref{eq:Damu} without violating other existing constraints.

A very
promising way to explain the deviation $\am{}$ is via scenarios beyond the SM (BSM)
with enhanced chirality flips. Such scenarios are also interesting
from the point of view of electroweak symmetry breaking
as they necessarily contribute to the fermion mass
generation mechanism and to effective Higgs-boson couplings \cite{Athron:2021iuf,Crivellin:2021rbq,Stockinger:2022ata,Dermisek:2023nhe}.
At the same time, there is currently no sign of new physics in
searches for charged lepton flavor violation (CLFV), despite the potential
of correlations of BSM effects on $\am$ and CLFV observables in
many concrete models. Typically, therefore, such models can only explain the deviation \eqref{eq:Damu} in non-generic parameter regions with large hierarchies between flavor-conserving and flavor-violating parameters. Here we study this conflict between $\am$ and CLFV in a concrete model, using the value of \eqref{eq:Damu} as an illustration. The conclusions of the present paper would essentially remain intact even if the deviation would reduce to a smaller value.

Leptoquark (LQ) models are among the best-motivated extensions of the
SM. Using the notation of Ref.\ \cite{Buchmuller:1986zs}, there are
two possible types of spin-0 LQ quantum numbers, $S_1$ and $R_2$,
which allow gauge invariant couplings to both left-handed and
right-handed leptons. These, therefore, allow enhanced chirality flips
and promising explanations of
$\am{}$ \cite{Chakraverty:2001yg,Mahanta:2001yc, Cheung:2001ip,Biggio:2014ela, Bauer:2015knc, Popov:2016fzr,Das:2016vkr,ColuccioLeskow:2016dox,Kowalska:2018ulj,Dorsner:2019itg,Crivellin:2020tsz,Gherardi:2020qhc,Babu:2020hun,Crivellin:2020mjs}. More generally, the $S_1$ and $R_2$ models are two of very few
viable single-field explanations of $\am{}$ \cite{Queiroz:2014zfa,Chiu:2014oma,Biggio:2014ela,Biggio:2016wyy,Athron:2021iuf}. In the past
years, LQ models have also frequently been proposed as combined
explanations of $B$-physics anomalies and $\am{}$ \cite{Bauer:2015knc,Das:2016vkr,Popov:2016fzr,Cai:2017wry,Crivellin:2017zlb,Nomura:2021oeu,Marzocca:2021azj,Zhang:2021dgl,Chen:2022hle,Freitas:2022gqs}, and models
with several leptoquarks are also able to simultaneously explain
neutrino masses \cite{Popov:2016fzr,Nomura:2021oeu,Freitas:2022gqs}. 
Refs.\ \cite{ColuccioLeskow:2016dox,Crivellin:2020tsz,Crivellin:2020mjs} confirm that the single LQ
explanations of $\am{}$ remain viable also given constraints on LQs from
Z-boson and Higgs-boson decays.

The $S_1$ and $R_2$ LQ models exemplify how large, chirality-flip
enhanced contributions to $\am{}$ can naturally be accompanied by
CLFV effects. Focusing on the $S_1$ model, its flavor structure is
governed by two $3\times3$ coupling matrices, i.e.\ by 18 free
parameters $\LRcouplings{q\ell}$ coupling left- or right-handed
quarks $q$ to leptons $\ell$. $\am{}$ depends on couplings of the
muon to the top- or charm-quark, while non-zero couplings of the electron and
$\tau$ lepton can lead to CLFV contributions.

Here we focus on the
impact of CLFV versus $\am{}$ constraints on the flavor 
structure of the $S_1$ LQ model. We aim for deriving general
constraints on the 18 flavor parameters, under the assumption that
the model explains $\am{}$.
To keep the analysis concrete, we consider several representative scenarios for the flavor structure which we call top-only, charm-only and up-type quark universal, as specified further in Sec.~\ref{sec:strategy}.
Our study is complementary to Ref.~\cite{Freitas:2022gqs}, where
$\am{}$, neutrino masses and complementary observables were fitted
to a minimal LQ model (containing $S_1$ and further particles and
general flavor coupling structure),
leading to specific best-fit values for the flavor 
parameters of the model. It is also complementary to
Refs.\ \cite{deBoer:2015boa,Mandal:2019gff}, where upper limits on
flavor parameters were derived without requiring an explanation of the
nonzero result for $\am{}$. In our case, the $S_1$ model alone cannot
explain neutrino masses; we aim for conservative and general limits
on the flavor parameters under the assumption that leptoquarks are
responsible for $\am{}$. The limits will be derived from correlations
between \am{} and various lepton flavor violation processes, such as
two-body decays \twodecays{}, three-body decays \threedecays{}, and
$\mu-e$ conversion in nuclei processes, as \mecold{} and \mec{}. To
manage a large number of free parameters, we restrict ourselves to
several specific cases, where the
anomalous magnetic moment of muon is explained either only by the top-quark or
by charm-quark contributions or by a combination thereof. 

The paper is structured as follows.
In Sec.~\ref{sec:definitions} we introduce our notations for the
$S_1$ leptoquark model, and Sec.~\ref{sec:analytical} with the
appendix present the
relevant analytical expressions of the considered observables.
Later, in Sec.~\ref{sec:strategy} the latest constraints on the
leptoquark mass from the LHC studies are shown and our analysis strategy is explained.
In Secs.~\ref{sec:am}\textendash{}\ref{sec:conversion} we show analytical results for observables under interest and derive the constraints on coupling constants that induce them.
Finally, the most important results are combined as conclusions in Sec.~\ref{sec:conclusions}.

\section{Model definition}\label{sec:definitions}
We consider the leptoquark $S_1$ model, which extends the SM particle
content by a single spin-0 leptoquark field with the gauge
representation $(\overline{\pmb{3}},\pmb{1},1/3)$ under the $SU(3)\times SU(2)\times U(1)$ group. 
The leptoquark is an $SU(2)$ singlet thus carrying an electric charge of $Q_{S_1} = 1/3$. 
The Lagrangian terms involving the $S_1$ leptoquark which are relevant for this study are expressed in the following way in the \emph{interaction} eigenstate basis (indicated by a tilde),
\begin{equation}
	\mathcal{L} \ni 
	- \ms{2} |S_1|^2 
	-\big(
	\tilde{\lambda}^{ql}_L~\overline{\tilde{Q}^c_q} i\sigma_2 \tilde{L}_{l}~S_1 
	+ \tilde{\lambda}^{ql}_{R}\overline{\tilde{u}_q^{Rc}}\tilde{e}^{R}_{l}S_1  + \hc{}
	\big)\,,
\end{equation}
containing the $SU(2)$-invariant product of the left-chiral quark and charged lepton doublet fields,
\begin{equation}
	\overline{\tilde{Q}^c_q} i\sigma_2 \tilde{L}_{l} = 
	\overline{\tilde u_q^{Lc}} \tilde e^L_l 
	- 
	\overline{\tilde d_q^{Lc}} \nu_l^L
	\,.
\end{equation}
This fermion-leptoquark interaction Lagrangian is the most
general one for the $S_1$ leptoquark type which prevents fast proton decay
by excluding couplings to quark-antiquark pairs. 

For  studying flavor physics, it is useful to rotate the fermion fields into mass eigenstates.
To perform this, the unitary matrices $U_{u,d,e}$, $V_{u,d,e}$ for
left- and right-handed fermion fields are applied (schematically as $\tilde\psi_k = U_{ik}^* \psi^{mass}_i \equiv
U_{ik}^* \psi_i$). 

In this way, the Standard Model Yukawa couplings and fermion mass
terms are diagonalized. The mixing matrices can be fully absorbed
in two out of the three leptoquark interaction terms with
left-/right-handed leptons and neutrinos. We choose the
so-called \emph{up-type mass diagonal basis}~\cite{Bigaran:2020jil,
	Dorsner:2020aaz}, where the new leptoquark coupling matrices are
defined as
\begin{align}
	\Lcoupling{ql} &= V^{\dagger iq}_u \tilde{\lambda}^{ij}_L V^{\dagger j
		l}_e\,,
	&
	\Rcoupling{ql} &= U^{\dagger iq}_d \tilde{\lambda}^{ij}_L U^{\dagger j
		l}_e\,.
\end{align}
Using these couplings the interaction Lagrangian contains interactions
with charged leptons and up-type quarks governed directly by the
$\LRcouplings{}$, while the interaction with neutrinos and down-type
quarks involves the CKM matrix $V_{\mathrm{CKM}}$,
\begin{equation}
	\mathcal{L} \ni 
	-\overline{u_q^c} 
	\big( \Lcoupling{ql} P_L + \Rcoupling{ql} P_R \big) e_{l} S_1 
	+ 
	\overline{d_q^c} \big( \Lcoupling{jl} V^{jq}_{\mathrm{CKM}} P_L \big) \nu_{l} S_1  + \hc{}
\end{equation}
As numerical values for the CKM matrix entries, we use the ones by the PDG~\cite{ParticleDataGroup:2022pth}.

\section{Analytical results}\label{sec:analytical}
In the present paper, we consider low-energy lepton observables as
constraints on the $S_1$ leptoquark model. The observables are the
muon magnetic moment $a_\mu=(g-2)_\mu/2$, two-body decays
\twodecays{}, three-body decays of the form \threedecays{}, and
$\mu\to e$ conversion in the presence of a
nucleus. Table~\ref{tab:observables-experiments} summarizes these
observables, current experimental limits, and expected sensitivities of
the next planned experiments. The present section collects analytical
results for the leptoquark contributions to all these
observables. Additional observables involving meson decays are
discussed in the appendix.

All one-loop results were obtained in two ways. First, by direct
Feynman diagrammatic calculation. Second, by automatic generation
using \texttt{FlexibleSUSY} \cite{Athron:2021kve,Athron:2014yba,Athron:2017fvs} and its extension package
\texttt{NPointFunctions}~\cite{Khasianevich:2022ess}.  
\texttt{FlexibleSUSY} is a \texttt{Mathematica} and
\texttt{C++} framework which compiles a spectrum generator out of a given
model definition input.  It uses \code{SARAH}~\cite{Staub:2015kfa,
	Vicente:2015zba}, for which we created a suitable model file
incorporating the parameterization of the couplings developed in
Sec.~\ref{sec:definitions}. 
This  setup  resulted in an independent cross-check of the
consistency of the results presented in the following. 

\subsection{\am{}}

\begin{figure}[t!]
	\begin{fmffile}{am}
		\def\subwidth{0.15\textwidth}
		\unitlength = 50pt
		\hspace*{\fill}
		\begin{subfigure}{\subwidth}
			\centering
			\vspace*{10pt}
			\begin{fmfgraph*}(1, 0.85)
				\fmfpen{0.5}\fmfstraight\fmfset{arrow_len}{2mm}
				\fmfleft{v2,v1}\fmfright{v4,v3}
				\fmf{phantom}{v2,v6,v4}
				\fmffreeze
				\fmf{fermion,tension=2,label=$\mu$,label.side=right}{v1,v5}
				\fmf{fermion,tension=2,label=$\mu$,label.side=right}{v7,v3}
				\fmf{boson,tension=2,label=$\gamma$}{v6,v8}
				\fmf{fermion,label=$q$}{v7,v5} 
				\fmf{dashes,label=$S_1$,label.side=right}{v5,v8} 
				\fmf{dashes,label=$S_1$,label.side=right}{v8,v7} 
			\end{fmfgraph*}
			\vspace*{10pt}
			\caption{SSF}
			\label{fig:am-ssf}
		\end{subfigure}
		\hfill
		\begin{subfigure}{\subwidth}
			\centering
			\vspace*{10pt}
			\begin{fmfgraph*}(1, 0.85)
				\fmfpen{0.5}\fmfstraight\fmfset{arrow_len}{2mm}
				\fmfleft{v2,v1}\fmfright{v4,v3}
				\fmf{phantom}{v2,v6,v4}
				\fmffreeze
				\fmf{fermion,tension=2}{v1,v5}
				\fmf{fermion,tension=2}{v7,v3}
				\fmf{boson,tension=2}{v6,v8}
				\fmf{dashes,label=$S_1$}{v7,v5} 
				\fmf{fermion,label=$q$,label.side=left}{v8,v5} 
				\fmf{fermion,label=$q$,label.side=left}{v7,v8} 
			\end{fmfgraph*}
			\vspace*{10pt}
			\caption{FFS}
			\label{fig:am-ffs}
		\end{subfigure}
		\hspace*{\fill}
	\end{fmffile}
	\caption{
		One-loop diagrams contributing to \am\ induced by $S_1$ leptoquark.
	}
	\label{fig:am-diagrams}
\end{figure}
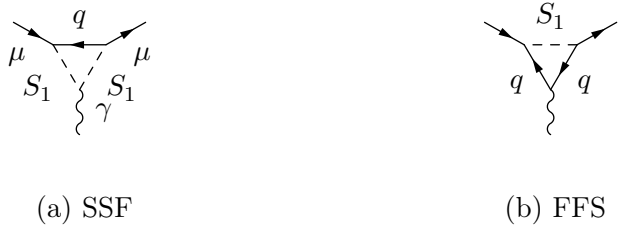

The two relevant one-loop Feynman diagrams contributing to \am,
i.e.\ the additional leptoquark contribution to $a_\mu$,
are depicted in Figure~\ref{fig:am-diagrams}.
Both diagrams have a very similar structure and involve an up-type
quark next to the leptoquark; they are often referred to as SSF  (see
Figure~\ref{fig:am-ssf}) and FFS (see Figure~\ref{fig:am-ffs}),
respectively.

Their sum can be written as
\begin{widetext}
\begin{equation}
	\begin{gathered}
		\Delta a^{\text{one-loop}}_{\mu} =
		\frac{m_{\mu}^2}{48 \pi^2 \ms{2}} 
		\bigg(
		\frac{m_{q}}{m_{\mu}} \Lcoupling{q2}\Rcoupling{q2}
		L_1(x_q) + \frac{\big(\Lcoupling{q2} \big)^2 + \big( \Rcoupling{q2} \big)^2}{4} 
		L_2(x_q)\bigg) 
		\,,\\
		\am  = 
		\deltaqed \Delta a^{\text{one-loop}}_{\mu},
		\qquad 
		\deltaqed = 1+\frac{e^2}{\pi^2}\ln\frac{m_\mu}{\ms{}}\,,
		\label{eq:am-full}
	\end{gathered}
\end{equation}
\end{widetext}
with the shorthand notation of the one-loop mass ratio argument $x_q =
m_{q}^2/\ms{2}$ used here and in the following. This result
coincides with the formulas presented e.g.\ in
Ref.~\cite{Athron:2021iuf} (see also references therein) and includes
universal leading logarithmic two-loop QED corrections
\deltaqed{}~\cite{Degrassi:1998es, vonWeitershausen:2010zr}, which are also implemented in
\texttt{FlexibleSUSY}, see Ref.~\cite{Athron:2017fvs}.
The loop function themselves are defined as (with following limits for $x\to0$: $F_F(x) \approx -9/2 - 3 \ln x$, $F_C(0)=0$, $F_E(0)=4$, $F_B(0)=2$; see also Refs.~\cite{Hisano:1995cp, Athron:2021iuf}):
\begin{widetext}
\begin{equation}
	\begin{alignedat}{4}
		L_1(x) &= 4 F_F(x) - F_C(x) > 0
		\,, \quad 
		&&L_2(x) &&= 2 F_E(x) - F_B(x) > 0 
		\,,\\
		F_F(x) &= \frac{3(-3+4x-x^2-2\ln x)}{2(1-x)^3}  \,,
		\quad
		&&
		F_E(x) &&= \frac{2(2+3x-6x^2+x^3+6x\ln x)}{(1-x)^4} \,,
		\\
		F_C(x) &= \frac{3(1-x^2+2x\ln x)}{(1-x)^3} \,,
		\quad
		&&F_B(x) &&= \frac{2(1-6x+3x^2+2x^3-6x^2\ln x)}{(1-x)^4} \,. 
		\label{eq:loop-functions-1}
	\end{alignedat}
\end{equation}
\end{widetext}
Note that the first two functions are positive, which allows only
constructive interference of contributions  from different quark
generations (as long as all couplings are positive).
The first term in Eq.~\eqref{eq:am-full} contains the chirally enhanced ratio $m_{q}/m_{\mu}$  whereas the second one does not.  
The chirally enhanced term appears together with a product of two couplings to different fermion chiralities $\Lcoupling{q2}\Rcoupling{q2}$.
It is well known that this enhancement is crucial for the possibility
to explain a significant leptoquark contribution to $\am{}$.

The theory prediction in Eq.~\eqref{eq:am-full} can be compared to the
difference between the experimental measurement and the corresponding
Standard Model prediction, see Eq.~\eqref{eq:Damu}. 

\subsection{\twodecays{}}

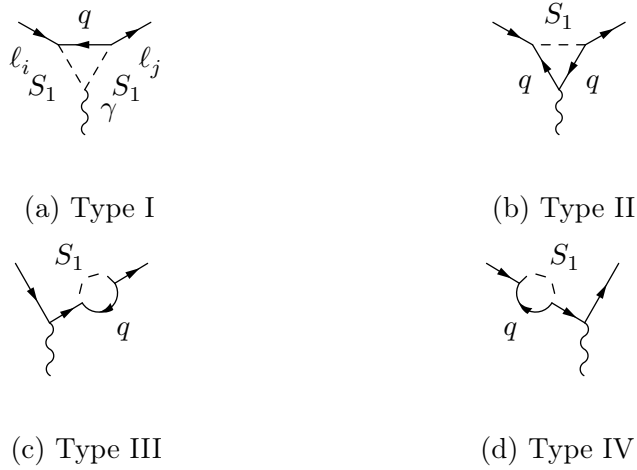
\begin{figure}[t!]
	\begin{fmffile}{meg}
		\def\subwidth{0.15\textwidth}
		\unitlength = 50pt
		\hspace*{\fill}
		\begin{subfigure}{\subwidth}
			\centering
			\vspace*{10pt}
			\begin{fmfgraph*}(1, 0.85)
				\fmfpen{0.5}\fmfstraight\fmfset{arrow_len}{2mm}
				\fmfleft{v2,v1}\fmfright{v4,v3}
				\fmf{phantom}{v2,v6,v4}
				\fmffreeze
				\fmf{fermion,tension=2,label=$\ell_i$,label.side=right}{v1,v5}
				\fmf{fermion,tension=2,label=$\ell_j$,label.side=right}{v7,v3}
				\fmf{boson,tension=2,label=$\gamma$}{v6,v8}
				\fmf{fermion,label=$q$}{v7,v5} 
				\fmf{dashes,label=$S_1$,label.side=right}{v5,v8} 
				\fmf{dashes,label=$S_1$,label.side=right}{v8,v7} 
			\end{fmfgraph*}
		\vspace*{10pt}
			\caption{Type I}
		\end{subfigure}
		\hfill
		\begin{subfigure}{\subwidth}
			\centering
			\vspace*{10pt}
			\begin{fmfgraph*}(1, 0.85)
				\fmfpen{0.5}\fmfstraight\fmfset{arrow_len}{2mm}
				\fmfleft{v2,v1}\fmfright{v4,v3}
				\fmf{phantom}{v2,v6,v4}
				\fmffreeze
				\fmf{fermion,tension=2}{v1,v5}
				\fmf{fermion,tension=2}{v7,v3}
				\fmf{boson,tension=2}{v6,v8}
				\fmf{dashes,label=$S_1$}{v7,v5} 
				\fmf{fermion,label=$q$,label.side=left}{v8,v5} 
				\fmf{fermion,label=$q$,label.side=left}{v7,v8} 
			\end{fmfgraph*}
			\vspace*{10pt}
			\caption{Type II}
		\end{subfigure}
		\hspace*{\fill}
		
		\hspace*{\fill}
		\begin{subfigure}{\subwidth}
			\centering
			\vspace*{10pt}
			\begin{fmfgraph*}(1, 0.85)
				\fmfpen{0.5}\fmfstraight\fmfset{arrow_len}{2mm}
				\fmfleft{v2,v1}\fmfright{v4,v3}
				\fmf{phantom, tension=2.8}{v2,v6}
				\fmf{phantom}{v6,v4}
				\fmffreeze
				\fmf{fermion}{v1,v5}
				\fmf{fermion}{v5,v8}
				\fmf{fermion}{v7,v3}
				\fmf{boson,tension=1.5}{v5,v6}
				\fmf{dashes,right,tension=0.01,label=$S_1$,label.dist=3}{v7,v8}
				\fmf{fermion,left,label=$q$}{v7,v8} 
			\end{fmfgraph*}
			\vspace*{10pt}
			\caption{Type III}
		\end{subfigure}
		\hfill
		\begin{subfigure}{\subwidth}
			\centering
			\vspace*{10pt}
			\begin{fmfgraph*}(1, 0.85)
				\fmfpen{0.5}\fmfstraight\fmfset{arrow_len}{2mm}
				\fmfleft{v2,v1}\fmfright{v4,v3}
				\fmf{phantom}{v2,v6}
				\fmf{phantom,tension=2.8}{v6,v4}
				\fmffreeze
				\fmf{fermion}{v1,v5}
				\fmf{fermion}{v8,v7}
				\fmf{fermion}{v7,v3}
				\fmf{boson, tension=1.5}{v6,v7}
				\fmf{dashes,left,tension=0.01,label=$S_1$,label.dist=3}{v5,v8}
				\fmf{fermion,left,label=$q$}{v8,v5} 
			\end{fmfgraph*}
			\vspace*{10pt}
			\caption{Type IV}
		\end{subfigure}
		\hspace*{\fill}
	\end{fmffile}
	\caption{One-loop diagrams contributing to \twodecays{} induced by $S_1$ leptoquark.}
	\label{fig:photon-diagrams}
\end{figure}

The Feynman diagrams contributing to two-body decays \twodecays{} are
similar to the ones contributing to
$\am{}$. Figure~\ref{fig:photon-diagrams} displays the four
contributing types of one-loop diagrams; the main difference is the
replacement of the external fermions and associated leptoquark
coupling constants.  

The contributions of Figure~\ref{fig:photon-diagrams} can be expressed
as amplitudes with off-shell photon with outgoing momentum $q = p_i -
p_j$ (using the conventions of
Refs.~\cite{Kotlarski:2019muo, Dudenas:2022von} with the covariant
derivative $\mathcal{D}_\mu=\partial_\mu + i e Q_f 
A_\mu$), 
\begin{equation}
	\begin{aligned}
		i\Gamma_{\bar \ell_j \ell_i \gamma}
		=\,&
		i \bar{u}_{j}
		\Big[
		\left(q^2 \gamma^\mu
		-q^\mu\slashed{q}\right)\left(A_1^{L} P_L + A_1^{R}P_R
		\right)
		+im_{i}\sigma^{\mu\nu}q_\nu \left(
		A_2^{L}P_L+A_2^{R}P_R\right)
		\Big] u_i
		\,.
		\label{2body-decay}
	\end{aligned}
\end{equation}
The two-body decays of interest only depend on the squares of the dipole form factors
$A_2^{L,R}$; the  branching ratio has the form (see,
e.g. \cite{Hisano:1995cp, Kotlarski:2019muo}):
\begin{equation}
	\begin{alignedat}{1}
		\br(\twodecays) & = 
		\frac{m_{\ell_i}^5}{16\pi\Gamma_{i}} \Big(|A_2^L|^2 + |A_2^R|^2\Big)
		\label{eq:two-full}
	\end{alignedat}
\end{equation}
with the decay width of muon and tau, $\Gamma_{\mu}=2.996 \cdot
10^{-19}~\gev$ and $\Gamma_{\tau}=2.267 \cdot
10^{-12}~\gev$~\cite{ParticleDataGroup:2020ssz}. 
The structure of the dipole form factors is analogous to the situation for \am{}, 
\begin{equation}
	\begin{aligned}
	A_2^L = 
		- \frac{1}{16\pi^2}\frac{e}{6\ms{2}}
		\Big(&
		\frac{m_q}{m_i}\Rcoupling{qj}\Lcoupling{qi} L_1(x_q)
		+
		\frac{1}{4}\Rcoupling{qj}\Rcoupling{qi} L_2(x_q)
		\Big) < 0
		\,.
	\end{aligned}	
\end{equation}
The expressions for $A_1$ terms will be listed below in the context of
three-body decays, where they will be relevant.

The prediction for the two-body decays will be compared to the
corresponding experimental upper limits listed in
Table~\ref{tab:observables-experiments}. The existing upper limits on
\meg{}, \teg{} and \tmg{} were obtained at
MEG~\cite{MEG:2020zxk}
BaBar~\cite{BaBar:2009hkt}, the next foreseeable
improvements are planned at 
MEG-II~\cite{MEGII:2018kmf} and
Belle-II~\cite{Banerjee:2022xuw}.

\subsection{\threedecays{}}

\begin{figure}[t!]
		\begin{fmffile}{meee}
		\def\subwidth{0.15\textwidth}
		\unitlength = 50pt
		\hspace*{\fill}
		\begin{subfigure}{\subwidth}
			\centering
			\vspace*{10pt}
			\begin{fmfgraph*}(1, 1)
				\fmfpen{0.5}\fmfstraight\fmfset{arrow_len}{2mm}
				\fmfleft{v2,v1}\fmfright{v4,v3}
				\fmf{fermion,tension=2,label=$\ell_i$,label.side=right}{v1,v5}
				\fmf{fermion,tension=2,label=$\ell_j$,label.side=right}{v7,v3}
				\fmf{fermion,tension=2,label=$\ell_k$}{v2,v6}
				\fmf{fermion,tension=2,label=$\ell_k$, label.side=right}{v6,v4}
				\fmf{boson,tension=2,label=$\gamma$}{v8,v6}
				\fmf{fermion,label=$q$}{v7,v5} 
				\fmf{dashes,label=$S_1$,label.side=right}{v5,v8} 
				\fmf{dashes,label=$S_1$,label.side=right}{v8,v7} 
			\end{fmfgraph*}
			\vspace*{10pt}
			\caption{Type I}
		\end{subfigure}
		\hfill
		\begin{subfigure}{\subwidth}
			\centering
			\vspace*{10pt}
			\begin{fmfgraph*}(1, 1)
				\fmfpen{0.5}\fmfstraight\fmfset{arrow_len}{2mm}
				\fmfleft{v2,v1}\fmfright{v4,v3}
				\fmf{fermion,tension=2}{v1,v5}
				\fmf{fermion,tension=2}{v7,v3}
				\fmf{fermion,tension=2}{v2,v6}
				\fmf{fermion,tension=2}{v6,v4}
				\fmf{boson,tension=2}{v6,v8}
				\fmf{dashes,label=$S_1$}{v7,v5} 
				\fmf{fermion,label=$q$,label.side=left}{v8,v5} 
				\fmf{fermion,label=$q$,label.side=left}{v7,v8} 
			\end{fmfgraph*}
			\vspace*{10pt}
			\caption{Type II}
		\end{subfigure}
		\hfill
		\begin{subfigure}{\subwidth}
			\centering
			\vspace*{10pt}
			\begin{fmfgraph*}(1, 1)
				\fmfpen{0.5}\fmfstraight\fmfset{arrow_len}{2mm}
				\fmfleft{v2,v1}\fmfright{v4,v3}
				\fmf{fermion}{v1,v5}
				\fmf{fermion}{v5,v8}
				\fmf{fermion,tension=2}{v2,v6,v4}
				\fmf{fermion}{v7,v3}
				\fmf{boson,tension=1.5}{v5,v6}
				\fmf{dashes,right,tension=0.01,label=$S_1$,label.dist=3}{v7,v8}
				\fmf{fermion,left,label=$q$}{v7,v8} 
			\end{fmfgraph*}
			\vspace*{10pt}
			\caption{Type III}
		\end{subfigure}
		\hspace*{\fill}
		
		\hspace*{\fill}
		\begin{subfigure}{\subwidth}
			\centering
			\vspace*{10pt}
			\begin{fmfgraph*}(1, 1)
				\fmfpen{0.5}\fmfstraight\fmfset{arrow_len}{2mm}
				\fmfleft{v2,v1}\fmfright{v4,v3}
				\fmf{fermion}{v1,v5}
				\fmf{fermion}{v8,v7}
				\fmf{fermion}{v7,v3}
				\fmf{fermion,tension=2}{v2,v6,v4}
				\fmf{boson,tension=1.5}{v6,v7}
				\fmf{dashes,left,tension=0.01,label=$S_1$,label.dist=3}{v5,v8}
				\fmf{fermion,left,label=$q$}{v8,v5} 
			\end{fmfgraph*}
			\vspace*{10pt}
			\caption{Type IV}
		\end{subfigure}
		\hfill
		\begin{subfigure}{\subwidth}
			\centering
			\vspace*{10pt}
			\begin{fmfgraph*}(1, 1)
				\fmfpen{0.5}\fmfstraight\fmfset{arrow_len}{2mm}
				\fmfleft{v2,v1}\fmfright{v4,v3}
				\fmf{fermion,tension=3}{v1,v5}
				\fmf{fermion,tension=3}{v2,v6}
				\fmf{fermion,tension=3}{v7,v3}
				\fmf{fermion,tension=3}{v8,v4}
				\fmf{fermion,tension=2,label=$q$,label.side=left}{v6,v5} 
				\fmf{dashes,tension=2,label=$S_1$,label.side=left}{v5,v7} 
				\fmf{fermion,tension=2,label=$q$,label.side=right}{v8,v7} 
				\fmf{dashes,tension=2,label=$S_1$,label.side=right}{v6,v8} 
			\end{fmfgraph*}
			\vspace*{10pt}
			\caption{Type V}
		\end{subfigure}
		\hspace*{\fill}
	\end{fmffile}
	\caption{
		One-loop diagrams contributing to $\br(\threedecays)$ induced by $S_1$ leptoquark.
		Diagrams of type I\textendash{}IV also have $u$-channel counterparts.
		In addition to the box diagram of type V there is an analogous one where leptoquarks propagate as quarks and vice versa.
		Higgs-boson penguins are negligible for the derivation of the LQ coupling limits due to SM Yukawa magnitude.
		$Z$-boson penguins lead to the contribution similar to the $A_1$ one of the photon but are relatively suppressed due to the mass of the former.
	}
	\label{fig:three-body-diagrams}
\end{figure}
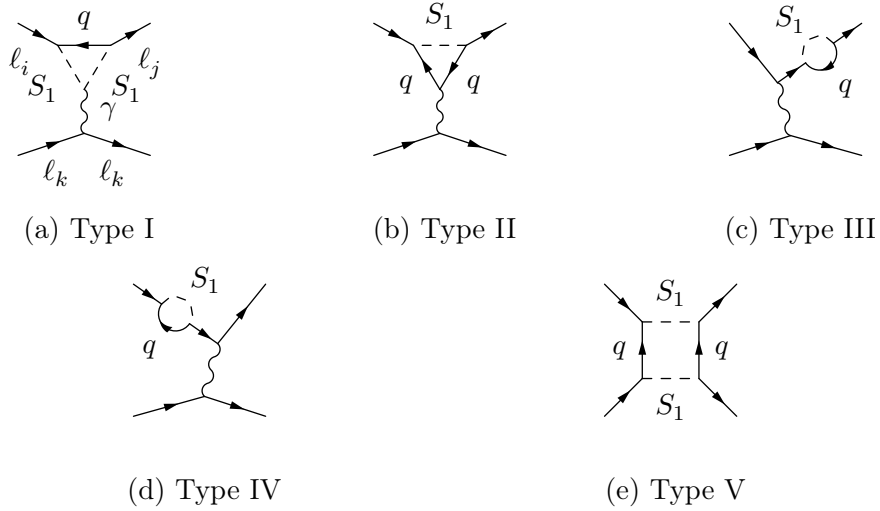

Like for $\am{}$ and for the two-body decays $\twodecays{}$,
leptoquarks contribute to the three-body decays $\threedecays{}$
starting from the one-loop level.
The five types of one-loop diagrams are shown in
Figure~\ref{fig:three-body-diagrams}.  Type I to IV contain a
\twodecays{} subdiagram but the outgoing on-shell photon is replaced by a virtual photon which finally decays into a lepton-antilepton pair. 
Type V is a box diagram that is
distinguished from all other diagrams in that it involves four powers
of leptoquark couplings instead of two.

The leptoquark contributions to the three-body decays arise via the dipole form
factor and 4-fermion (scalar, vector, and tensor) form factors
$S_{XY}$, $V_{XY}$, $T_{XY}$ (with $X,Y$ being $L$ or $R$).
The
vector form factor receives a contribution not only from actual
4-fermion box diagrams of type V in
Figure~\ref{fig:three-body-diagrams}, denoted as $V_{XY}^\Box$, but
also from the photonic form 
factor $A_1$ defined above in Eq.~\eqref{2body-decay}.

The full form of the decay width for \threedecays{} in case of $j=k$ and $j\ne k$ reads, see also Refs.~\cite{Okada:1999zk,Crivellin:2017rmk,Kuno:1999jp,Ilakovac:1994kj}:
\begin{widetext}
\begin{equation}
	\begin{alignedat}{5}
		\Gamma_{\ell_i\to3\ell_j} = \frac{m_i^5}{192\pi^3}\Big(
		&e^2 |A_2^L|^2\big(\ln\tfrac{m_i^2}{m_j^2}-\tfrac{11}{4}\big)
		+
		e\big(
		\tfrac{3}{2}eA_1^L-\tfrac{1}{2}(V_{LL}^{\Box}+V_{LR}^{\Box})\big)|A_2^R|
		\\
		+&
		\tfrac{1}{4}V_{LL}^2 + \tfrac{1}{8}V_{LR}^2 + \tfrac{1}{16}S_{LL}^2 
		+[L\leftrightarrow R]
		\Big)
		\,,
		\label{eq:three-gamma-same}
	\end{alignedat}
\end{equation}
\begin{equation}
	\begin{alignedat}{5}
		\Gamma_{\threedecays} = \frac{m_i^5}{192\pi^3}
		\Big(
		&e^2 |A_2^L|^2 \big(\ln\tfrac{m_i^2}{m_k^2}-3\big)
		+
		e\big(
		eA_1^L-\tfrac{1}{2}(V_{LL}^{\Box}+V_{LR}^{\Box})\big)|A_2^R|
		\\
		+&
		\tfrac{1}{8}\big(V_{LL}^2+V_{LR}^2\big)
		+
		\tfrac{1}{32}\big(S_{LL}^2+S_{LR}^2\big)
		+
		\tfrac{3}{2}T_{LL}^2
		+[L\leftrightarrow R]
		\Big)
		\,.
		\label{eq:three-gamma-different}
	\end{alignedat}
\end{equation}
\end{widetext}
The vectorial photon form factor $A_1$ and its contribution to the
4-fermion form factors are given by
\begin{equation}
	\begin{alignedat}{2}
		A_1^L& = \frac{1}{16\pi^2}\frac{e}{36\ms{2}}L_3(x_q)\Lcoupling{qj}\Lcoupling{qi}
		\,,
		\\
		V_{XY} &=
		- e A^X_1 + nV_{XY}^\Box
		\,,
	\end{alignedat}
\end{equation}
where the minus sign is related to the form factor embedding into the
4-fermion amplitude, and where a similar equation holds for
$A_1^R$; $n=\frac{1}{2}$ for $V_{XX}$ in $\ell_i\to3\ell_j$, and $n=1$ otherwise. The loop function takes the form (with
	the following limit for $x\to0$: $F_A(0) = 0$, $F_D(x)\approx 4 (4 +
	3 \ln x)$): 
\begin{equation}
	\begin{gathered}
		L_3(x) = F_A(x) - 2F_D(x) > 0
		\,, \\
		F_A(x) = \frac{
			2 - 9x + 18x^2 - 11x^3 + 6 x^3 \ln x
			}{(1-x)^4}
		\,,\\
		F_D(x) = \frac{
			16 - 45 x + 36 x^2 - 7 x^3 + 6 (2-3x)\ln x
			}{(1-x)^4}
		\,.
	\end{gathered}
\end{equation}

The pure box diagram contributions can be written as
\begin{widetext}
\begin{equation}
	\begin{alignedat}{5}
		S_{LL} &= \frac{1}{16\pi^2}\big(
		\tfrac{1}{2}
		\Lcoupling{q_2i} \Lcoupling{q_1k}
		-
		\Lcoupling{q_1i} \Lcoupling{q_2k}
		\big)
		\Rcoupling{q_1j} \Rcoupling{q_2k}
		m_{q_1}m_{q_2}D_0
		\,,\\
		S_{LR} &= -\frac{1}{16\pi^2}\big(
		2\Lcoupling{q_2i} \Rcoupling{q_1k}
		|D_{00}|
		+
		\Lcoupling{q_1i} \Rcoupling{q_2k}
		m_{q_1}m_{q_2}D_0
		\big)\Rcoupling{q_1j} \Lcoupling{q_2k}
		\,,\\
		V^{\Box}_{LL} &= \frac{1}{16\pi^2}\big(
		\Lcoupling{q_1i} \Lcoupling{q_2k}
		+
		\Lcoupling{q_2i} \Lcoupling{q_1k}
		\big)
		\Lcoupling{q_1j} \Lcoupling{q_2k}
		|D_{00}|
		\,,\\
		V^{\Box}_{LR} &= \frac{1}{16\pi^2} \big(
		\Lcoupling{q_1i} \Rcoupling{q_2k}
		|D_{00}| 
		+
		\Lcoupling{q_2i} \Rcoupling{q_1k}
		\tfrac{1}{2}m_{q_1}m_{q_2}D_0
		\big)
		\Lcoupling{q_1j} \Rcoupling{q_2k}
		\,,\\
		T_{LL} &= \frac{1}{16\pi^2}
		\Rcoupling{q_1j} \Rcoupling{q_2k} \Lcoupling{q_2i} \Lcoupling{q_1k}
		\tfrac{1}{8}m_{q_1}m_{q_2}D_0 
		\label{eq:all-box-coefficients}
	\end{alignedat}
\end{equation}
\end{widetext}
with expressions for $S_{RL}, S_{RR}, V^{\Box}_{RL}, V^{\Box}_{RR}, T_{RR}$ obtained by replacing $L\leftrightarrow R$; the zero-momenta Passarino-Veltman coefficient functions $D_0$ and $D_{00}$~\cite{Passarino:1978jh} can be simplified to for $q=c~\text{or}~t$:
\begin{equation}
	\begin{alignedat}{3}
		&\ms{4} D_0\big|_{q_1=q_2} = 
		\frac{-2 + 2 x_q - (1 + x_q) \ln x_q}{(1 - x_q)^3} > 0
		\,,\\
		&\ms{2} D_{00}\big|_{q_1=q_2} = 
		-\frac{1 - x_q^2 + 2 x_q \ln x_q}{4(1 - x_q)^3} < 0
		\,,\\
		&\ms{4} D_0\big|_{m_c\to0} =
		\frac{-1 + x_t - \ln x_t}{(1 - x_t)^2}>0
		\,,
		\\&
		\ms{2} D_{00}\big|_{m_c\to0} = 
		\frac{-1 + x_t - x_t \ln x_t}{4 (1 - x_t)^2} < 0
		\,.
	\end{alignedat}
\end{equation}

The prediction for the three-body decays will be compared to the
corresponding experimental upper limits listed in
Table~\ref{tab:observables-experiments}. Most important are the limits
on \meee. The existing limits were obtained at		SINDRUM~\cite{SINDRUM:1987nra}, and the next foreseen improvement is planned at Mu3e-I~\cite{Mu3e:2020gyw}.

We note that the main contribution from the box factors comes from the
$D_{00}$-terms and that the box form factors
in~\eqref{eq:all-box-coefficients} are positive, for positive values
of the $\LRcouplings{}$.

\subsection{$\mu\to e$ conversion}

\begin{figure}[t!]
	\begin{fmffile}{mec}
		\def\subwidth{0.2\textwidth}
		\unitlength = 50pt
		\hspace*{\fill}
		\begin{subfigure}{\subwidth}
			\centering
			\vspace*{10pt}
			\begin{fmfgraph*}(1, 0.75)
				\fmfpen{0.5}\fmfstraight\fmfset{arrow_len}{2mm}
				\fmfleft{v2,v1}\fmfright{v4,v3}
				\fmf{fermion,label=$\mu$,label.side=left}{v1,v5}
				\fmf{fermion,label=$u_{i}$,label.side=right}{v2,v5}
				\fmf{fermion,label=$e$,label.side=left}{v6,v3}
				\fmf{fermion,label=$u_j$,label.side=right}{v6,v4}
				\fmf{dashes,label=$S_1$}{v5,v6}
			\end{fmfgraph*}
			\vspace*{10pt}
		\end{subfigure}
		\hspace*{\fill}
	\end{fmffile}
	\caption{Tree-level diagram contributing to \mecold{} and \mec{}.}
	\label{fig:lonely-tree}
\end{figure}
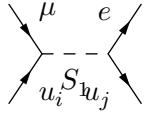

Out of all considered observables $\mu\to e $ conversion in presence
of a nucleus is special since it is mediated already by a tree-level diagram with
leptoquark exchange. Figure~\ref{fig:lonely-tree} shows the
diagram. The resulting predicted conversion rate can be  expressed as
\begin{equation}
	\br(\mu-e) = \frac{	\big(\alpha_{s} \Rcoupling{12} - \alpha_{v} \Lcoupling{12}\big)^2
		\big(\Lcoupling{11}\big)^2
		+ [L\leftrightarrow R]}{4\ms{4}\omega_{\mathrm{capt}}} 
	\label{eq:mec-overall}
\end{equation}
with the capture rate and the form-factors (muon mass and $\omega_{\mathrm{capt}}$ are expressed in \gev{} units):
\begin{equation}\label{eq:mec-formfactors}
	\begin{alignedat}{3}
		&\alpha_{s}=\sum_{i=p,n}f^i_u\tfrac{m_i}{m_u} S^{(i)}
		&&\!=
		1.537~(0.430~\text{in }Al) \cdot m_{\mu}^{5/2}
		\,,\\
		&\alpha_{v}=2V^{(p)}+V^{(n)}
		&&\!=
		0.280~(0.049~\text{in }Al) \cdot m_{\mu}^{5/2}
		\,,\\
		&\omega_{\mathrm{capt}}
		&&\!=
		8.849~(0.464~\text{in }Al) \cdot 10^{-18}\,.
	\end{alignedat}
\end{equation}
The overlap integrals are taken from the second method in Ref.~\cite{Kitano:2002mt}.
The proton and neutron scalar couplings $f^{p,n}_u$ are determined from pion-nucleon $\sigma_{\pi N}$ term for $u$ quark (see the Ref.~\cite{Hoferichter:2015dsa} for the numerical values).
Vector form-factors 2 and 1 (in $\alpha_v$) do not suffer from
theoretical uncertainty and are derived from the conservation of
vector current consideration, i.e. counting of valence quarks. 

The past SINDRUM-II and the planned COMET-I (as well as COMET-II and
Mu2e~\cite{Bernstein:2019fyh}) experiments listed in 
Table~\ref{tab:observables-experiments} use either $Au$ or $Al$ nuclei
for $\mu\to e $ conversion.
The prediction can be applied to both cases, see the first/second
numbers in Eq.~\eqref{eq:mec-formfactors} accordingly. 

\section{Analysis strategy}\label{sec:strategy}
Our main interest is the impact of $\am{}$ and CLFV observables on the
full $3\times3$ coupling matrices $\LRcouplings{}$, using the
experimental bounds listed in Table~\ref{tab:observables-experiments}.
To simplify the analysis, we assume all 18 considered couplings to be positive  and apply the customary perturbative upper bound~\cite{Allwicher:2021rtd, Bandyopadhyay:2021kue}:
\begin{equation}
	0 < \Lcoupling{ij} < \sqrt{4\pi}
	\,,\qquad
	0 < \Rcoupling{ij} < \sqrt{4\pi}
	\,,
	\label{eq:coupling-limits}
\end{equation}
on each matrix element.

\begin{table*}[t!]
	\centering
	\begin{tabular}{| c | c | c | c |}
		\hline
		Observable & 
		Current phase & 
		Next phase\\
		\hline	
		$\Delta a_\mu^{2021}$ & 
		FNAL~\cite{Muong-2:2021ojo}: $(25.1\pm5.9)\cdot 10^{-10}$ &
		\textemdash{}
		\\\hline
		\meg &
		MEG~\cite{MEG:2020zxk}: $4.2\cdot 10^{-13}$
		&
		MEG-II~\cite{MEGII:2018kmf}: $6\cdot 10^{-14}$
		\\
		\teg & 
		BaBar~\cite{BaBar:2009hkt}: $3.3\cdot10^{-8}$
		&
		Belle-II~\cite{Banerjee:2022xuw}: $9.0\cdot10^{-9}$
		\\
		\tmg &
		BaBar~\cite{BaBar:2009hkt}: $4.4\cdot10^{-8}$
		&
		Belle-II~\cite{Banerjee:2022xuw}: $6.9\cdot10^{-9}$
		\\
		\hline
		\meee
		& 
		SINDRUM~\cite{SINDRUM:1987nra}: $1\cdot 10^{-12}$
		&
		Mu3e-I~\cite{Mu3e:2020gyw}: $2\cdot 10^{-15}$
		\\
		\teee &
		Belle-I~\cite{Hayasaka:2010np}: $2.7\cdot10^{-8}$
		&
		Belle-II~\cite{Banerjee:2022xuw}: $4.7\cdot10^{-10}$
		\\
		\tmee &
		Belle-I~\cite{Hayasaka:2010np}: $1.8\cdot10^{-8}$
		&
		Belle-II~\cite{Banerjee:2022xuw}: $2.9\cdot10^{-10}$
		\\
		\temm&
		Belle-I~\cite{Hayasaka:2010np}: $2.7\cdot10^{-8}$
		&
		Belle-II~\cite{Banerjee:2022xuw}: $4.5\cdot10^{-10}$
		\\
		\tmmm& 
		Belle-I~\cite{Hayasaka:2010np}: $2.1\cdot10^{-8}$
		&
		Belle-II~\cite{Banerjee:2022xuw}: $3.6\cdot10^{-10}$
		\\
		\hline
		\mecold &
		SINDRUM-II~\cite{SINDRUMII:2006dvw}: $7\cdot10^{-13}$&
		\textemdash{}
		\\
		\mec &
		\textemdash{}
		&
		COMET-I~\cite{COMET:2018auw}: $7\cdot 10^{-15}$
		\\
		\hline
		\kplus &
		E949~\cite{E949:2008btt}: $1.73\cdot 10^{-10}$ &
		\textemdash{}
		\\
		\dzero &
		LHCb~\cite{LHCb:2013jyo}: $7.6\cdot 10^{-9}$ &
		\textemdash{}
		\\
		\hline
	\end{tabular}
	\caption{
		Experimental bounds, that are considered in
		the present paper. The column ``Current phase'' refers to
		current, existing bounds, and the column ``Next phase'' refers to
		the next available expected future bound. We use 90 \%
		confidence level (but $1\sigma$-bound in case of \am). 
		{Note, the anomalous magnetic moment of muon and kaon branching ratio are the only quantities corresponding to observations and not upper limits.}
	} 
	\label{tab:observables-experiments}	
\end{table*}

\begin{table*}[t!]
	\centering
	\begin{tabular}{| c | c | c | c |}
		\hline
		Decay/coupling & $\beta$ & Lowest allowed mass [\gev{}] & Reference	\\
		\hline
		${ue}$ & 1.0 (0.5) & 1435 (1270) & $\sqrt{s}=13~\tev$ CMS~\cite{CMS:2018ncu} \\
		${ue}$ & 1.0 (0.5) & 1400 (1290) & $\sqrt{s}=13~\tev$ ATLAS~\cite{ATLAS:2019ebv} \\
		$\lambda^{ue}=1.0~(0.8)$ & 1.0 & 1755 (1355) & $\sqrt{s}=8~\tev$ CMS~\cite{CMS:2015xzc} \\
		\hline
		${c\mu}$ & 1.0 (0.5) & 1530 (1285) & $\sqrt{s}=13~\tev$ CMS~\cite{CMS:2018lab} \\
		${c\mu}$ & 1.0 (0.5) & 1560 (1230) & $\sqrt{s}=13~\tev$ ATLAS~\cite{ATLAS:2019ebv}\\
		$\lambda^{c\mu}=1.0$ & 1.0 & 660 & $\sqrt{s}=8~\tev$ CMS~\cite{CMS:2015xzc} \\
		\hline
		${t\mu}$ & 1.0 & 1420 & $\sqrt{s}=13~\tev$ CMS~\cite{CMS:2018oaj} \\
		\hline
		${t\tau}$ & 1.0 & 950 & $\sqrt{s}=13~\tev$ CMS~\cite{CMS:2020wzx} \\
		${t\tau}$ & 1.0 (0.5) & 920 (810) & $\sqrt{s}=13~\tev$ ATLAS~\cite{ATLAS:2019qpq} \\
		$\lambda^{t\tau}=2.5$ & 1.0 & 1020 & $\sqrt{s}=13~\tev$ CMS~\cite{CMS:2020wzx}	\\
		\hline
	\end{tabular}
	\caption{
		LHC constraints on scalar leptoquarks masses at 95\% confidence level. 
		The first column shows the decay mode assumed in the analysis,
		or --- for analyses considering single leptoquark production --- specifies the assumed value of the relevant coupling. In
		the second column the quantity $\beta$ is the leptoquark
		branching decay ratio into the quark/lepton
		mentioned in the first column. The numbers without brackets
		correspond to the strongest achievable bounds, the numbers in
		brackets correspond to alternative assumptions and
		corresponding weaker bounds.
	}
	\label{tab:lhc-limits}
\end{table*}

Possible masses of leptoquarks are constrained by a variety of LHC
analyses accumulated in Table~\ref{tab:lhc-limits}. 
In this paper, we fix the leptoquark mass in all numerical results below as
\begin{equation}
	\ms{} = 1.8~\tev\,.
\end{equation}  
This value is conservative as it respects all current LHC restrictions in the third column of Table~\ref{tab:lhc-limits}.

In our analysis, we focus particularly on three distinct
scenarios.
This helps manage the 18-dimensional parameter space and
draw illuminating and fairly general conclusions. 
The leptonic
observables mainly correlate the coupling matrices $\LRcouplings{q\ell}$ horizontally --- i.e.\
couplings of the same quark to different leptons. 
	This is different from the case of e.g.\ $B$-physics and the
	constraints from accommodating $B$-anomalies related to $R(D^{(*)})$
	as done e.g.\ in Ref.\ \cite{Crivellin:2017zlb}.
For this reason, our scenarios leave this horizontal direction
unconstrained but impose various vertical relationships on the
coupling matrices.

\emph{Scenario 1, top-only case}: Here only couplings to the top-quark
are nonzero. We are left with the 6 parameters $\LRcouplings{3\ell}$,
$\ell=1,2,3$.

\emph{Scenario 2, charm-only case}: Here only couplings to the charm-quark
are nonzero. We are left with the 6 parameters $\LRcouplings{2\ell}$,
$\ell=1,2,3$.

\emph{Scenario 3, columns case}: Here we assume quark-universality of
couplings, i.e.\ assume equal couplings in each column of the coupling
matrices,
$\Lcoupling{1\ell}=\Lcoupling{2\ell}=\Lcoupling{3\ell}\equiv\Lcoupling{\ell}$ (and
the same for \Rcoupling{}). We are left with the 6 parameters $\LRcouplings{\ell}$,
$\ell=1,2,3$.

In addition, we will use the $\mu$--$e$ conversion process to
constrain the up-quark couplings $\LRcouplings{1\ell}$ ($\ell=1,2$) in
a way independent of assumptions on vertical relationships. 
In all analyses, we will only consider real couplings.
We note that similar but more restrictive scenarios were discussed in Refs.~\cite{Cheung:2001ip, Chakraverty:2001yg, Biggio:2014ela, Kowalska:2018ulj,Dorsner:2019itg, Crivellin:2020tsz, Bigaran:2020jil} to study muon $g-2$, and similar scenarios allowing for CLFV processes were discussed in Refs.~\cite{Mahanta:2001yc, Bauer:2015knc,  Popov:2016fzr,Das:2016vkr,ColuccioLeskow:2016dox,Gherardi:2020qhc,Babu:2020hun,Crivellin:2020mjs, Dorsner:2020aaz}.

\section{Phenomenological consequences of \am{}}\label{sec:am}
\begin{figure*}[t!]
	\hspace*{\fill}
	\begin{subfigure}[c]{\picwidth}
		\includegraphics[width=\textwidth]{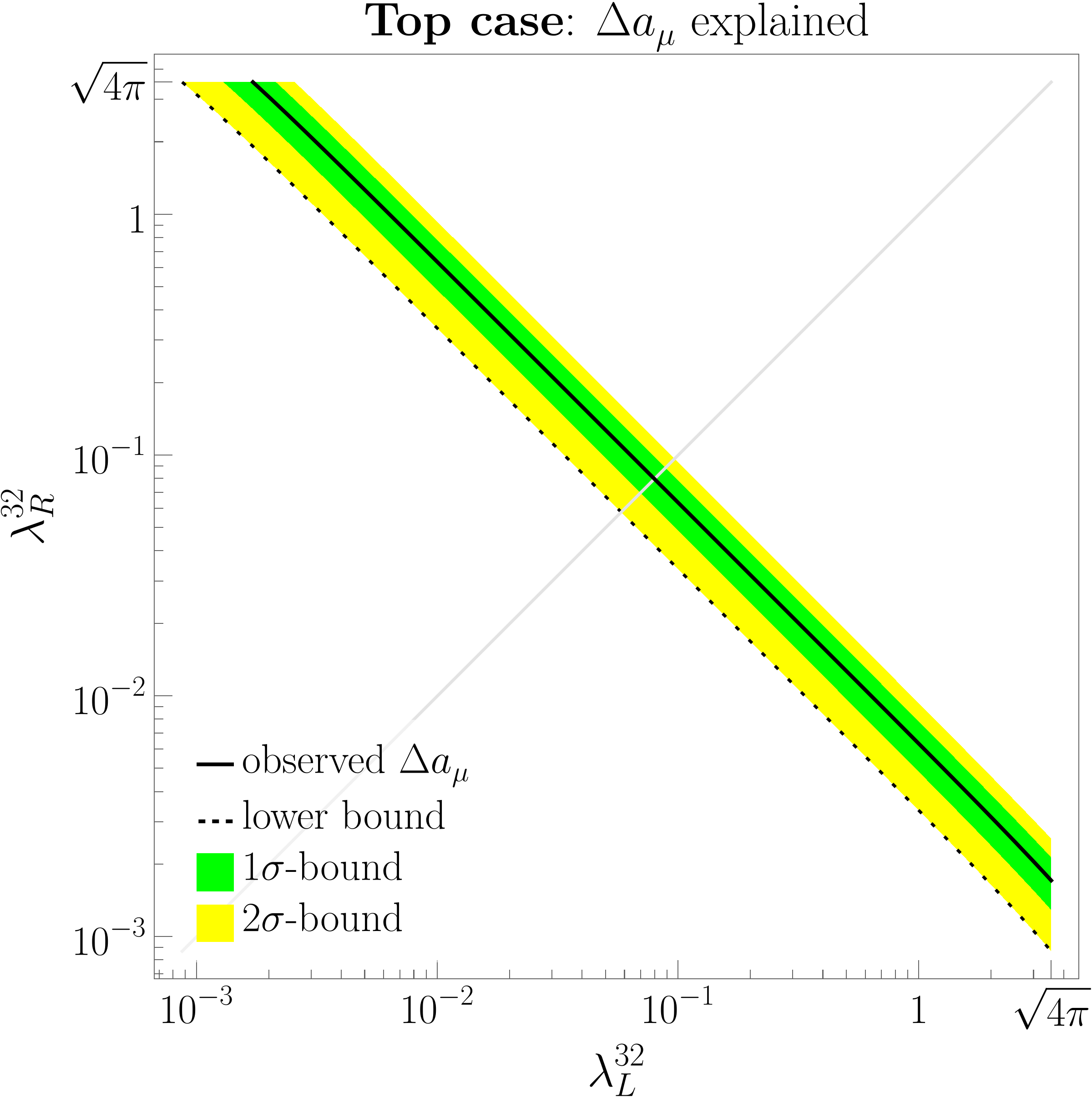}
		\caption{}
		\label{fig:am-top}
	\end{subfigure}
	\hfill
	\begin{subfigure}[c]{\picwidth}
		\includegraphics[width=\textwidth]{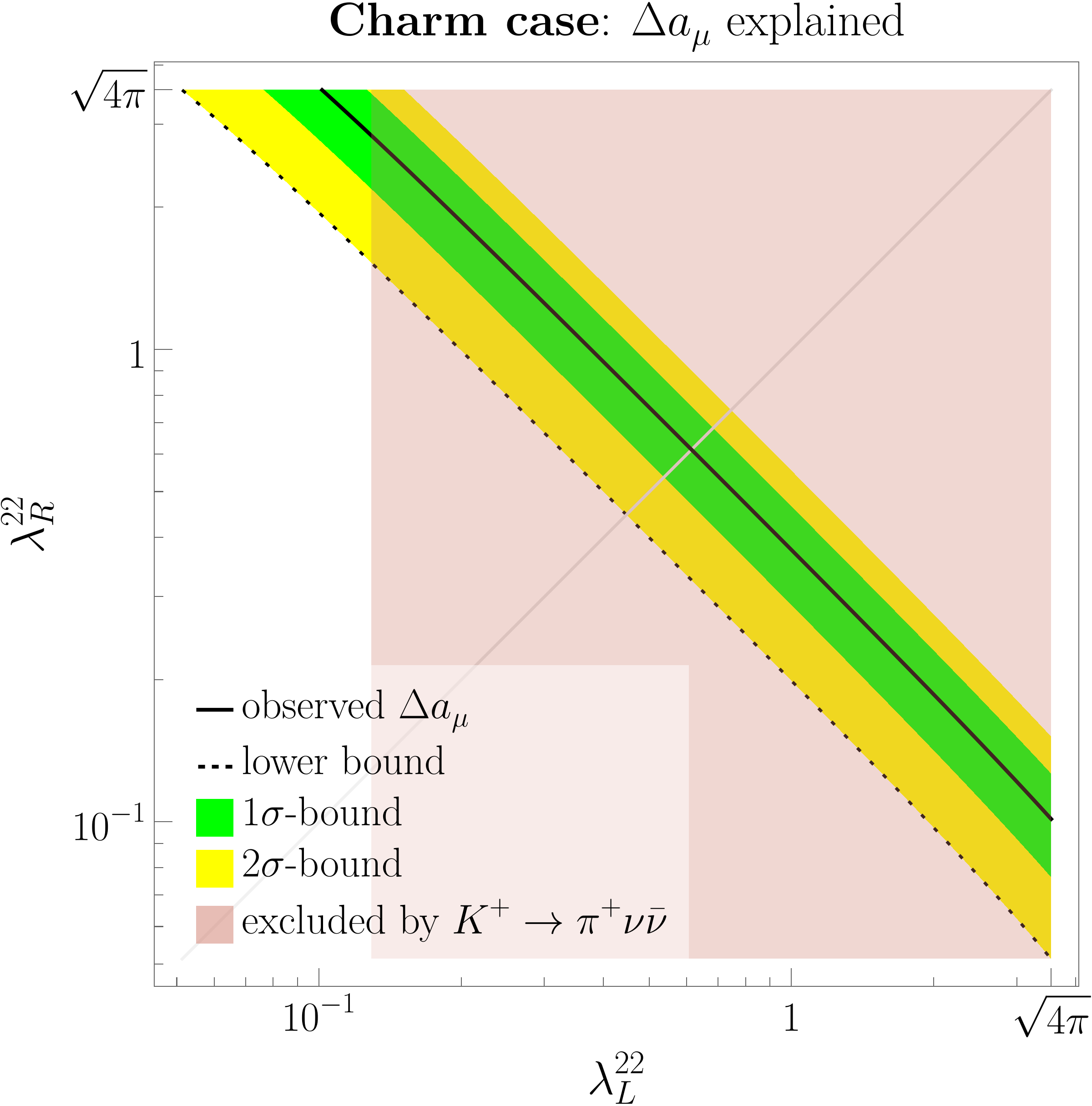}
		\caption{}
		\label{fig:am-charm}
	\end{subfigure}
	\hspace*{\fill}
	
	\begin{subfigure}[c]{\picwidth}
		\includegraphics[width=\textwidth]{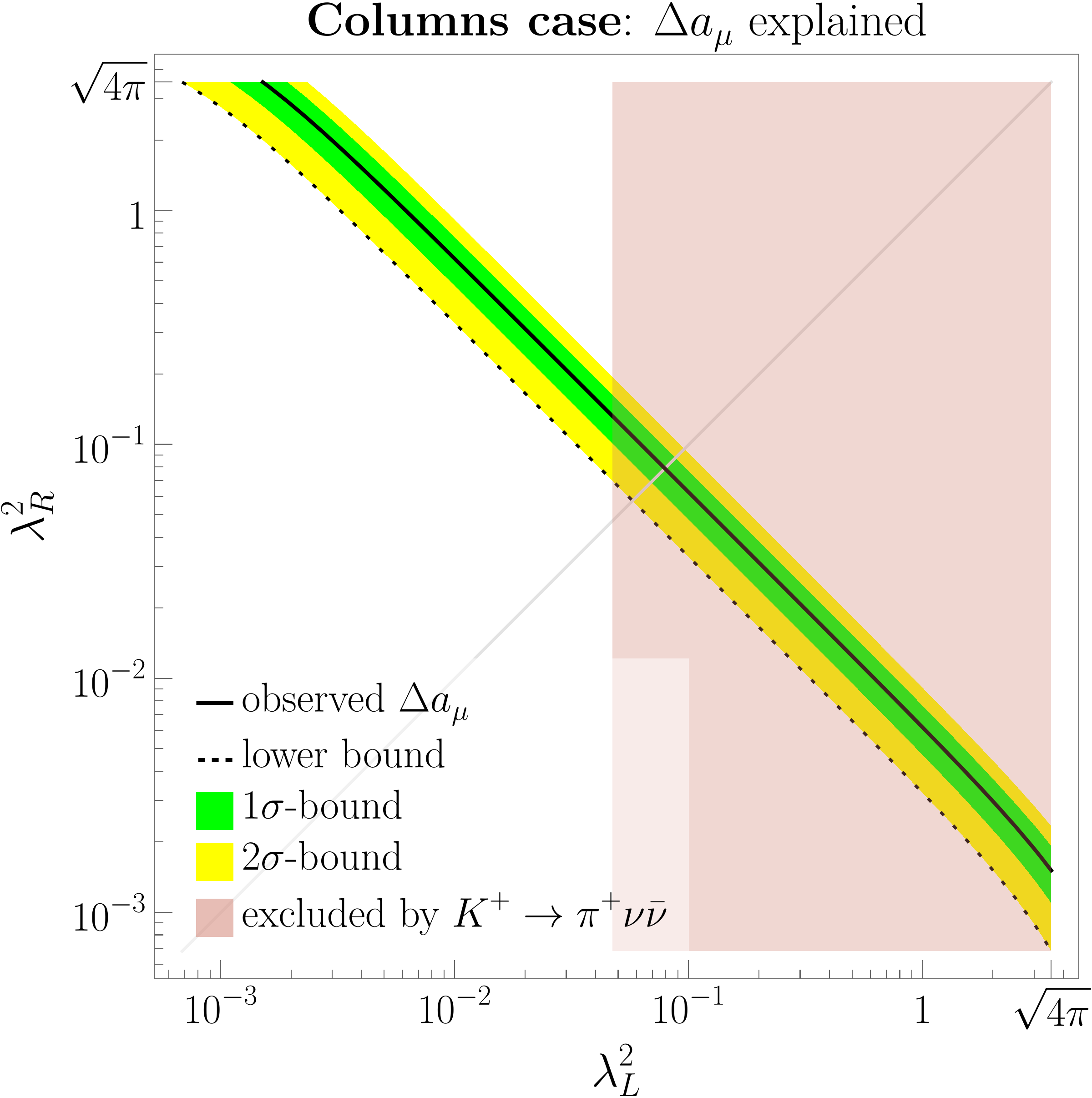}
		\caption{}
		\label{fig:am-columns}
	\end{subfigure}
	\caption{
		Bands in parameter space allowed by \am{},  for
		$\ms{}=1.8~\tev{}$. The three plots correspond to the
		three different scenarios, defined in Sec.~\ref{sec:strategy}. The red-shaded regions correspond
		to the maximal additional limits from the \kplus{}
		decay.
	}\label{fig:am}
\end{figure*}

We begin our phenomenological investigations with an analysis of
\am{}. In addition to known results in the literature (see
in particular Refs.\ \cite{Kowalska:2018ulj,Bigaran:2020jil,Athron:2021iuf}) we focus
on the contributions of all generations and derive bounds on several
(combinations of) $\LRcouplings{}$-parameters which will be instructive and
useful later.

The analytical result was presented in Eq.\ \eqref{eq:am-full}. It
contains chirality-flipping terms proportional to
$m_q\Lcoupling{q2}\Rcoupling{q2}$ where $q$ is one of the up-type
quarks. It is well known that leptoquark models can explain large
\am{} only via such chirality-flipping terms which are enhanced by the
large top- or charm-quark masses.

To provide an overview we first record a criterion under which
chirality-flip enhancement is at all possible. In
Eq.~\eqref{eq:am-full} the relative factors between the
chirality-flipping and non-flipping terms are schematically
$m_q\Lcoupling{q2}\Rcoupling{q2}L_1:m_\mu |\LRcouplings{q2}|^2L_2$ 
with loop functions $L_{1,2}$. 
Inserting typical masses of the order few TeV we obtain restrictions on the ratios between the left- and
right-handed couplings corresponding to chiral enhancement:
	
\begin{equation}\label{eq:chiralenhancementcriterion}
	\begin{alignedat}{3}
		\text{charm:}\quad&&\frac{1}{700}&\lesssim
		\frac{\Lcoupling{22}}{\Rcoupling{22}}\lesssim700 
		\,,\\
		\text{top:}\quad&&
		\frac{1}{4\cdot10^4}&\lesssim
		\frac{\Lcoupling{32}}{\Rcoupling{32}}\lesssim4\cdot10^4
		\,.
	\end{alignedat}
\end{equation}

\noindent
In Eq.~\eqref{eq:chiralenhancementcriterion} we have not included the
up-quark, since its contributions to \am{} are generally small.
Indeed, the \kplus{} decay, see Figure~\ref{fig:kaon-solo}, implies the
restriction $\Lcoupling{12}\lesssim 1$
regardless of all other relevant couplings due to cancellation in $\hat Y^L_{12}$ factor, see appendix~\ref{app:kplus}.
With the upper perturbative limit applied for the other coupling
$\Rcoupling{12}\lesssim \sqrt{4\pi}$, the maximum contribution of the
first quark-generation couplings contributes at most $9\%$ of the \am{} mean value (for $\ms{}=1.8~\tev$).
This number falls drastically for heavier leptoquark mass and/or
smaller couplings.
Hence, one can to a very good extent neglect the up-quark contributions
and focus on the ones from heavier quarks.


Now we focus on the first of our scenarios and consider the
\emph{top-couplings} \LRcouplings{32} and their values required to
explain \am{}, while the charm-/up-quark couplings are set to zero.

Due to $m_{t}/m_{\mu} \sim O(10^3)$ the full prediction for \am{} of
Eq.~\eqref{eq:am-full}
can be well approximated (if the
coupling ratio is in the range \eqref{eq:chiralenhancementcriterion})
by the chirality-flipping term, which in turn can be approximated as (\emph{top-only scenario})
\begin{equation}
	\am
	\approx
	3.3 \cdot 10^{-7}
	~
	\frac{1+0.64 \ln(\ms{}/2\,\tev)}{(\ms{}/2\,\tev)^2}
	~
	\Lcoupling{32}\Rcoupling{32}
	\,,
	\label{eq:top-am-product}
\end{equation}
which highlights the dependence on the couplings and allows to read
off easily the values for masses in the few-TeV range.

From comparing with the experimental result we get  bounds on products
of the two couplings that are approximately located within hyperbolic curves in the
$\Lcoupling{32}-\Rcoupling{32}$ plane. This is shown in the double logarithmic
scale in Figure~\ref{fig:am-top}, where the hyperbolic shape becomes a
straight band. The plot shows the coupling values for which the
experimental \am{} result is explained at the $1\sigma$ ($2\sigma$)
level in green (yellow). The plot is obtained from the exact
Eq.~\eqref{eq:am-full}, hence there is an ${\cal O}(10\%)$ distortion
from the hyperbolic shape due to the non-chirally enhanced terms.

Despite the small distortion, the band in Figure~\ref{fig:am-top}
essentially restricts the product of the left- and right-handed
top-couplings to the muon. As a simple formula, the entire $2\sigma$
band is confined in the interval  (\emph{top-only scenario})
\begin{equation}
	\begin{alignedat}{2}
		\text{\am{} band:}&\quad
		3.1 \cdot 10^{-3}
		< \Lcoupling{32} \Rcoupling{32} < 
		9.3 \cdot 10^{-3}
		\,.
		\label{eq:top-am-lower-and-upper-limit}
	\end{alignedat}
\end{equation} 

If we apply the perturbativity upper limit $\sqrt{4\pi}$ on each
individual coupling, the product~\eqref{eq:top-am-lower-and-upper-limit} implies also lower limits on
each coupling  (\emph{top-only scenario}):
\begin{equation}
	\text{2$\sigma$ individual limit:}\quad
	8.7 \cdot 10^{-4}
	< \LRcouplings{32}
	\,.
	\label{eq:individual-top-am}
\end{equation}
Note that the $2\sigma$ label here does not have a direct statistical
meaning but refers to the $2\sigma$ bound of Figure~\ref{fig:am-top}
from which the limit is derived.

As a by-product, this equation also implies a possible range of the
ratio of couplings $  {\Lcoupling{32}}:{\Rcoupling{32}} $ between
around $1/4000\ldots4000$, which is a sharpened version of
Eq.~\eqref{eq:chiralenhancementcriterion} derived only from
chirality-flip dominance.


Now we repeat the analysis for the second scenario and consider
explaining \am{} purely with the \emph{charm-couplings}
\LRcouplings{22}, setting the top-/up-quark couplings to zero. 
The ratio $m_{c}/m_{\mu} \sim O(10)$ is smaller than the one for the top-quark.
Nevertheless, it makes the chirally enhanced term still dominate such
that the non-chirally enhanced term can be neglected to estimate how
strongly the relevant couplings are restricted.

Applying similar simplifications as in the top case one obtains
the following approximation which
highlights the dependence on the couplings and is valid in the few-TeV
range (\emph{charm-only scenario}): 
\begin{equation}
	\am 
	\approx
	5.4 \times 10^{-9}
	~
	\frac{1+0.14 \ln(\ms{}/2\,\tev)}{(\ms{}/2\,\tev)^2}
	~
	\Lcoupling{22}\Rcoupling{22}
	\,.
	\label{eq:charm-am-product}
\end{equation}
Figure~\ref{fig:am-charm} shows the corresponding bands in the
$\Lcoupling{22}-\Rcoupling{22}$ plane explaining the measured \am{} at
the $1\sigma$ and $2\sigma$ level. The distortion of the hyperbolic
shape is stronger compared to the top-quark case because the dominance of the chirality-flipping contributions is less
pronounced. Still, it is essentially the product of the two couplings
which matters for \am{}, and it is again meaningful to provide the
interval of the coupling product for the entire $2\sigma$ band (\emph{charm-only scenario}):
\begin{equation}
	\begin{alignedat}{2}
		\text{\am{} band:}&\quad
		0.18 < 
		\Lcoupling{22} \Rcoupling{22} 
		<
		0.56
		\,.
		\label{eq:charm-am-lower-and-upper-limit}
	\end{alignedat}
\end{equation} 

As shown in Ref.~\cite{Kowalska:2018ulj}, there is a bound from the
measurement of $\br(\kplus)$ which imposes an additional restriction
on the coupling \Lcoupling{22}.\footnote{%
In addition, Drell-Yan dilepton processes $pp\to \mu^+\mu^-$ and $pp\to \mu^+\mu^-j$ provide an upper allowed value for \Rcoupling{22} as function of $\Lcoupling{22}$ and \ms{}, see Ref.~\cite{Kowalska:2018ulj,Raj:2016aky, Bansal:2018eha, Fuentes-Martin:2020lea}.
This upper bound excludes part of the $2\sigma$ bound for the charm-only scenario in Eq.~\eqref{eq:charm-am-lower-and-upper-limit}.
We do not use this upper bound here, because for us, it is the lower bound on the couplings in Eq.~\eqref{eq:charm-am-lower-and-upper-limit}, which impacts the analyses in the remainder of the paper.}
However, the bound significantly depends on \Lcoupling{12}: for lower
values we obtain the result from \cite{Kowalska:2018ulj}, for larger
ones \Lcoupling{22} becomes unrestricted.
This behavior is illustrated by the upper boundary of the
green area in the Figure~\ref{fig:kaon-solo} of the appendix~\ref{app:kplus}.
If one sets the coupling  \Lcoupling{12} to zero, which implies the
maximally restrictive bound from the \kplus\ decay, then the excluded
region is shown in Figure~\ref{fig:am-charm} as the pink area. 

Similarly to the top-case, applying the perturbativity upper limit on
each coupling together with
Eq.~\eqref{eq:charm-am-lower-and-upper-limit}  leads to lower limits
on each coupling. These lower limits, and the additional
limits from the \kplus\ decay (which applies for the specific case where
\Lcoupling{12} vanishes) can be summarized as (\emph{charm-only scenario})
\begin{equation}
	\label{charmindividuallimits}
	\begin{alignedat}{3}
		\text{2$\sigma$ individual limit:}&\quad
		5.1 \cdot 10^{-2}
		< \LRcouplings{22}
		\,,\\
		\text{\kplus{} limits:}&\quad
		\Lcoupling{22}< 0.13,~1.5<\Rcoupling{22}
		\,.
	\end{alignedat}
\end{equation}


Finally, we focus on the third scenario, the \emph{columns case} where
the leptoquark couplings are universal over the quark generations.
In this case, \am{} is dominated by top-quark contributions and the
bounds on the universal couplings are similar to the ones in the
top-only case, however the additional limits from the \kplus{} decay
are driven by a combination of up- and charm-quark couplings. The
corresponding plot is shown in Figure~\ref{fig:am-columns}, and the
limits are (\emph{columns scenario})
\begin{equation}
	\begin{alignedat}{2}
		\text{\am{} band:}&\quad
		2.4 \cdot 10^{-3}
		< \Lcoupling{2} \Rcoupling{2} < 
		9.2 \cdot 10^{-3}
		\,,\\
		\text{2$\sigma$ individual
			limit:}&\quad
		6.8 \cdot 10^{-4}
		< \LRcouplings{2}
		\,,\\
		\text{\kplus{} limits:}&\quad
		\Lcoupling{2} < 4.7\cdot10^{-2},~7.0\cdot10^{-2}<\Rcoupling{2}
		\,.
	\end{alignedat}
\end{equation} 

\section{Phenomenological consequences of two-body decays \meg, \teg, and \tmg}\label{sec:two-body}
\subsection{Consequences of decays involving muons}

Now we consider the impact of CLFV on the leptoquark couplings,
with special focus on the condition that the current \am{} is
explained. The first set of CLFV observables are the decays $\mu\to
e\gamma$ and $\tau\to\mu\gamma$. These have the common feature that,
like \am{},
they involve the muon and are governed by a dipole interaction which
can be dominated by chirality-flipping terms.

We begin with the analysis of the top-related couplings \Lcoupling{3i}
and \Rcoupling{3i} in scenario 1 (see Sec.~\ref{sec:strategy}), where the up- and charm-related
couplings are assumed to vanish. Like for \am{}, see
Eq.~\eqref{eq:top-am-product}, an instructive approximation is
obtained by taking only the chirally enhanced terms in the
formula~\eqref{eq:two-full} for the decays \twodecays. In this
approximation, the limits on branching ratios from
Table~\ref{tab:observables-experiments}  translate into the following
inequality (all masses are to be given in units of \gev{}; \emph{top-only scenario}):
\begin{equation}
	\norm{ \Rcoupling{3i} \Lcoupling{3j} } +
	\norm{ \Lcoupling{3i} \Rcoupling{3j} } < 
	\frac{\Gamma_i\br(\twodecays)}{m_i^3}
	\frac{0.73\ms{4}}{(1-0.17 \ln\ms{})^2}
	\,.
	\label{eq:two-top-overall} 
\end{equation}
For fixed $i,j$, this is a limit on a combination of four couplings.
There are several ways to extract more detailed information on bounds.

First we may fix the couplings \LRcouplings{32} relevant for \am{}
such that the experimental \am{} is explained, i.e.\ fix a point in
the band of Figure~\ref{fig:am-top}. In this way, two out of the four
couplings are fixed, and e.g.\ for $\mu\to e\gamma$,
Eq.~\eqref{eq:two-top-overall} takes the structure
$a|\Lcoupling{31}|^2+b|\Rcoupling{31}|^2<c$, i.e.\ it restricts the
remaining two couplings onto an ellipse.

\begin{figure*}[t!]
	\hspace*{\fill}
	\begin{subfigure}[c]{\picwidth}
		\includegraphics[width=\textwidth]{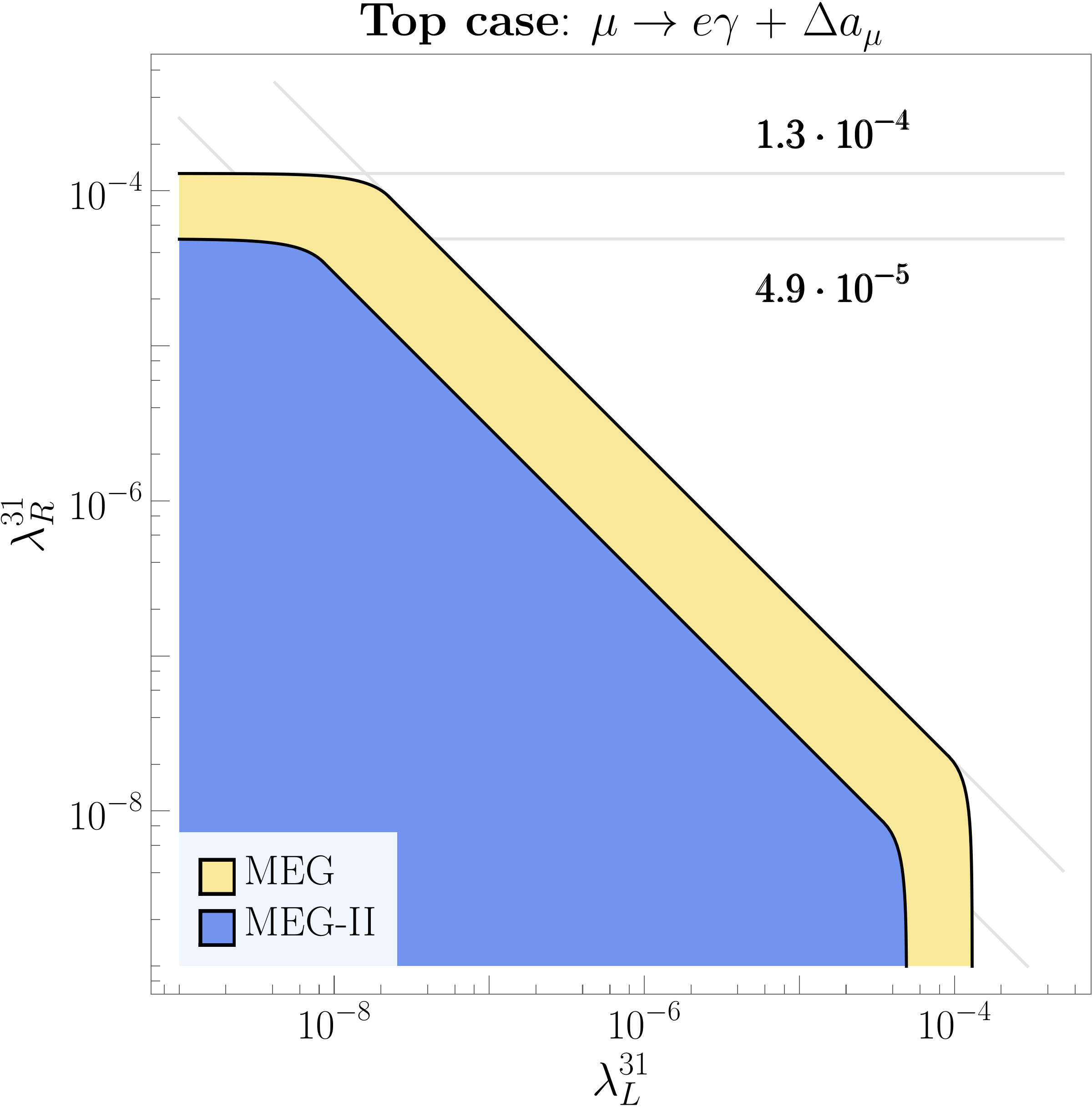}
		\caption{}
		\label{fig:top-meg}
	\end{subfigure}
	\hfill
	\begin{subfigure}[c]{\picwidth}
		\includegraphics[width=\textwidth]{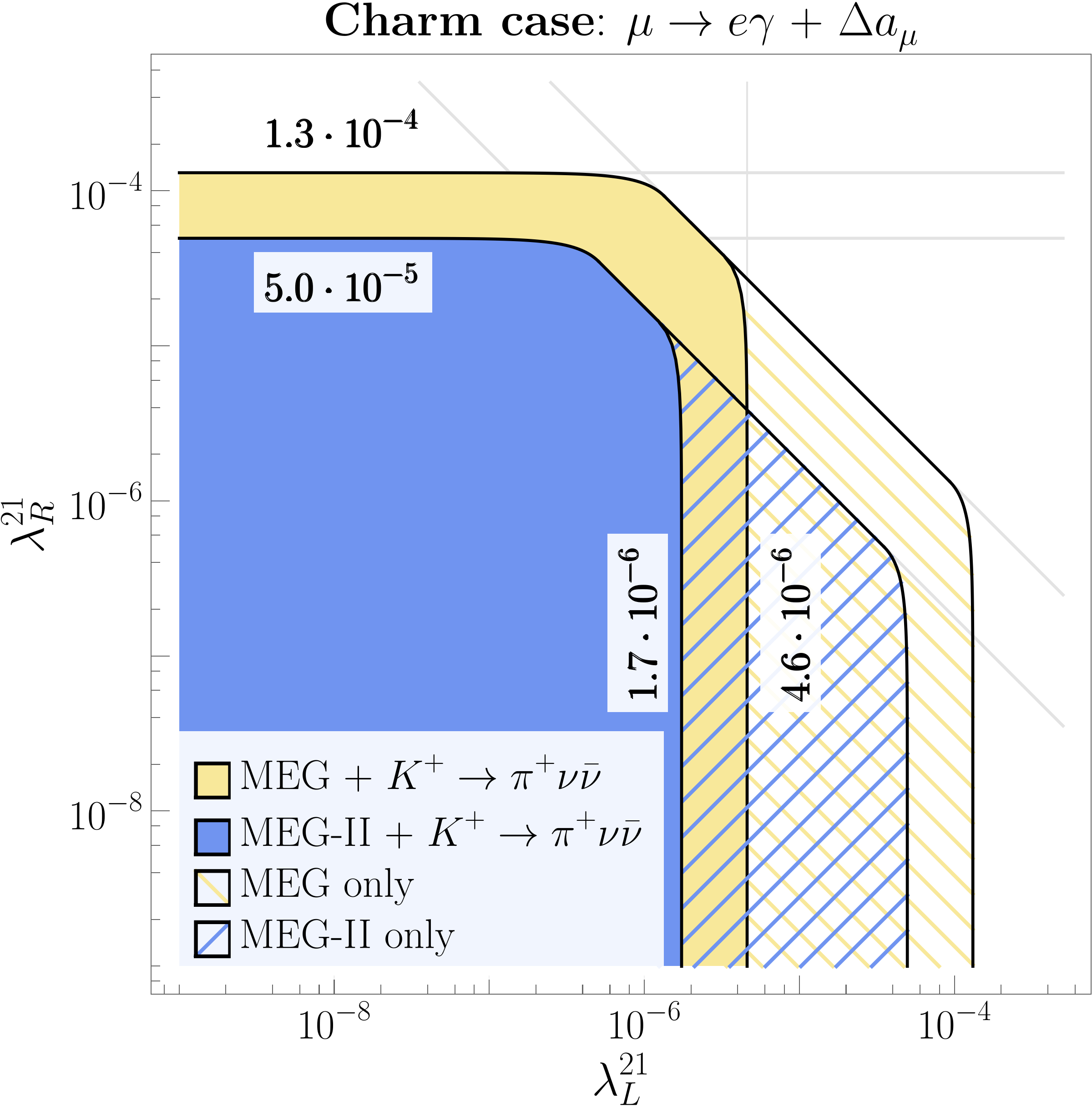}
		\caption{}
		\label{fig:charm-meg}
	\end{subfigure}
	\hspace*{\fill}

	\begin{subfigure}[c]{\picwidth}
		\includegraphics[width=\textwidth]{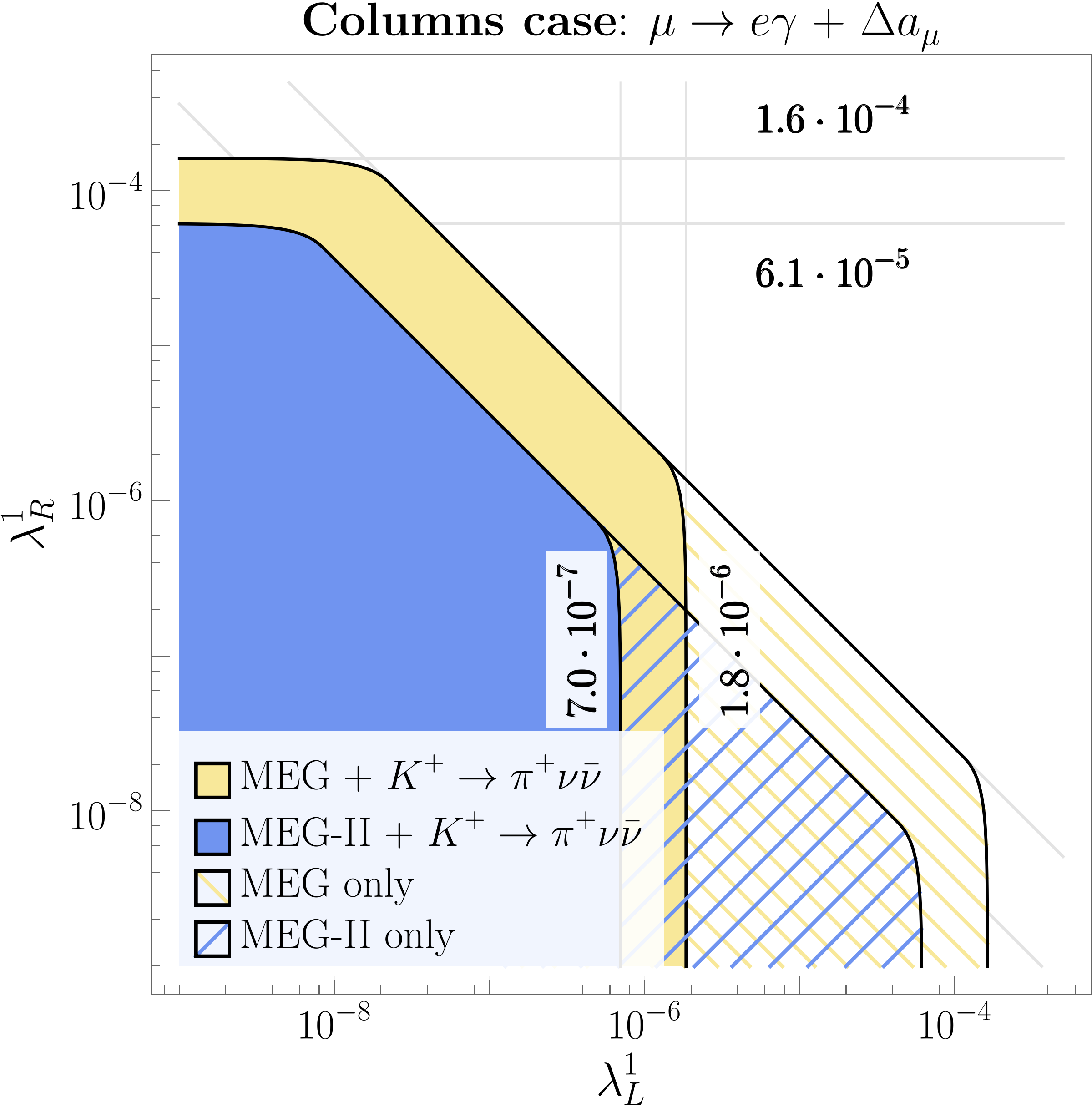}
		\caption{}
		\label{fig:column-meg}
	\end{subfigure}
	\caption{Allowed parameter regions for the \meg{} decay.
		assuming that \am{} is explained, with $\ms{}=1.8~\tev$, and
		for the  three different scenarios defined in Sec.~\ref{sec:strategy}. The meaning of the additional
		limits from the \kplus{} decay is as in
		Figure~\ref{fig:am}. 
	}
	\label{fig:meg}
\end{figure*}

\begin{figure*}[t!]
	\hspace*{\fill}
	\begin{subfigure}[c]{\picwidth}
		\includegraphics[width=\textwidth]{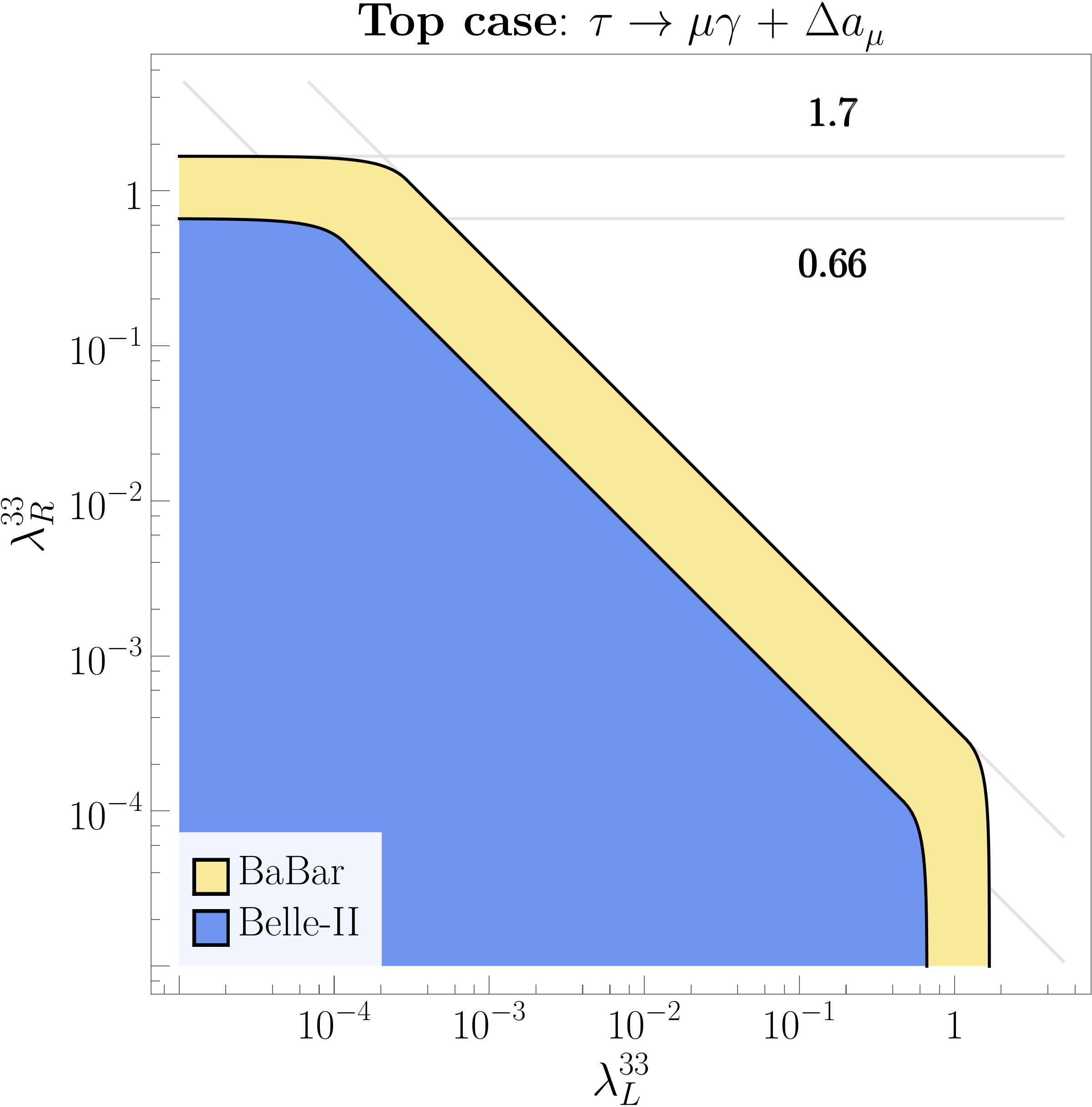}
		\caption{}
		\label{fig:top-tmg}
	\end{subfigure}
	\hfill
	\begin{subfigure}[c]{\picwidth}
		\includegraphics[width=\textwidth]{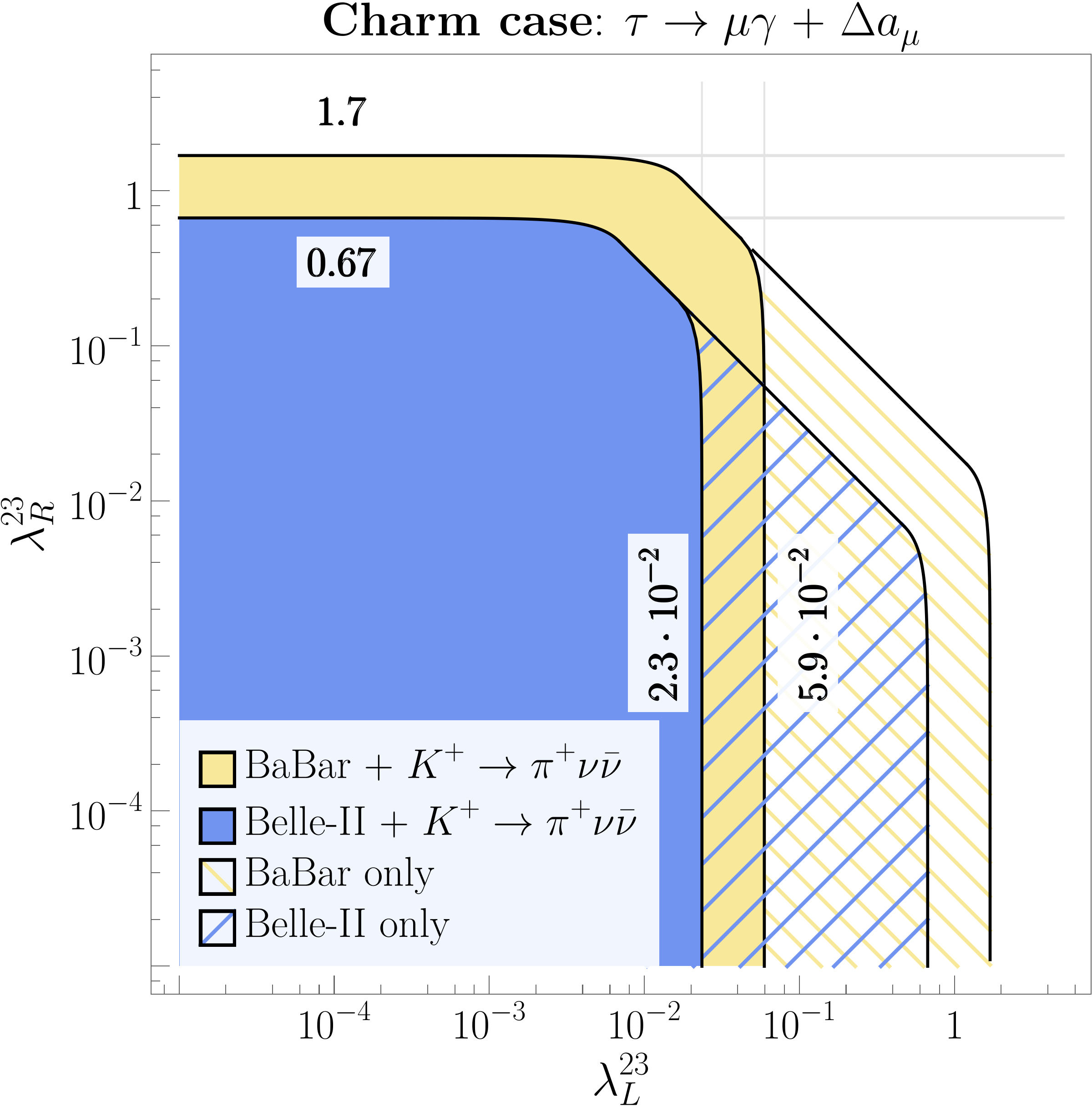}
		\caption{}
		\label{fig:charm-tmg}
	\end{subfigure}
	\hspace*{\fill}
	
	\begin{subfigure}[c]{\picwidth}
		\includegraphics[width=\textwidth]{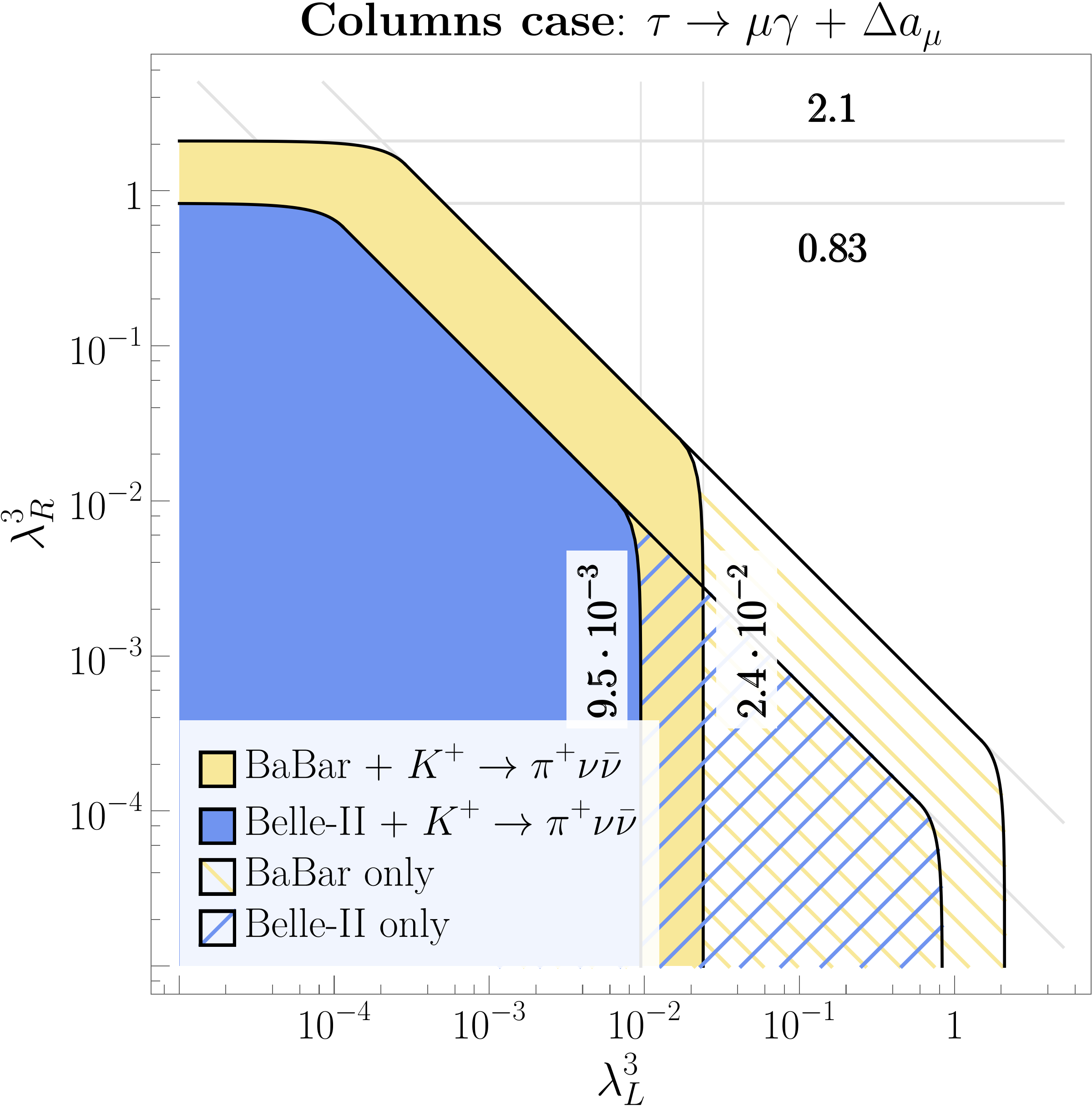}
		\caption{}
		\label{fig:column-tmg}
	\end{subfigure}
	\caption{As Figure~\ref{fig:meg} but for the \tmg{} decay.
	}
\end{figure*}

It turns out that the unification of all such ellipses is essentially
a hyperbolic region. 
This observation allows to decouple the influence of the \am{}-related
couplings from  consideration.
Figure~\ref{fig:top-meg}  shows
the corresponding allowed parameter regions in the plane of the two
couplings  \Lcoupling{31} and \Rcoupling{31}. It is obtained from a
scan over parameters (for $\ms{}=1.8~\tev$), requiring that 
\Lcoupling{32} and \Rcoupling{32}  are chosen such that the $\am{}$ 
prediction is within a $2\sigma$ band around the measured value quoted
in Table \ref{tab:observables-experiments}.
The yellow (blue) regions are allowed by the bounds of the ``current
phase'' (``next phase'') experiments in
Table~\ref{tab:observables-experiments}. Figure~\ref{fig:top-tmg} is analogous but
for the decay $\tau\to\mu\gamma$ and for the couplings   \Lcoupling{33} and \Rcoupling{33}.

To explain the shape of the allowed regions, we at first introduce
auxiliary variables that are the ratio of left and right couplings: 
\begin{equation}
	\ratio{ij} = \frac{\Lcoupling{ij}}{\Rcoupling{ij}}
	\,,
	\label{eq:ratio-definition}
\end{equation}
and rewrite the limit in Eq.~\eqref{eq:two-top-overall} for
$\ms{}=1.8~\tev$ equivalently as (\emph{top-only scenario})
\begin{equation}
	\begin{alignedat}{2}
		\Lcoupling{31} \Rcoupling{31}\Lcoupling{32} \Rcoupling{32}
		&< 
		3.0\cdot10^{-2}
		~&&\br(\meg) \frac{\ratio{31}\ratio{32}}{k_{31}^2+k_{32}^2}
		\,,\\
		\Lcoupling{33} \Rcoupling{33}\Lcoupling{32} \Rcoupling{32}
		&< 
		4.8\cdot10
		~&& \br(\tmg) \frac{\ratio{33}\ratio{32}}{k_{33}^2+k_{32}^2}
		\,.
		\label{eq:top-two-kk}
	\end{alignedat}
\end{equation}
The  \ratio{ij}-dependent factor is maximal for equal \ratio{ij} ratios
and together with the minimal \am{}-allowed product of $32$-couplings,
see  Eq.~\eqref{eq:top-am-lower-and-upper-limit}, provides the
most conservative (in the case of purely top-related couplings) bounds
on the product of couplings. These bounds take the announced
hyperbolic shape, i.e.\ they depend only on the products of two
couplings (\emph{top-only scenario}): 
\begin{equation}
	\begin{alignedat}{4}
		&\meg\big|_{\am}:~
		&&\Lcoupling{31} \Rcoupling{31}
		&&< 
		2.1\cdot10^{-12}
		&&\rightarrow 
		2.9\cdot10^{-13}
		\,,\\
		&\tmg\big|_{\am}:~
		&&\Lcoupling{33} \Rcoupling{33} 
		&&< 
		3.5\cdot10^{-4}
		&&\rightarrow 
		5.4\cdot10^{-5}
		\,.
		\label{eq:top-am-connected-limits}
	\end{alignedat}
\end{equation}
Here and in the following the first (second) number on the right-hand sides correspond to the ``current
phase'' (``next phase'') experiments and the yellow (blue) regions in
Figures~\ref{fig:top-meg} and ~\ref{fig:top-tmg}. In the figures (with logarithmic
scale) these hyperbolic limits are
visible as the inclined lines.

Figures~\ref{fig:top-meg} and ~\ref{fig:top-tmg} also show that the hyperbolic shape is cut
off by individual upper limits on each coupling (\emph{top-only scenario}):
\begin{equation}
	\label{eq:individualtwobody}
	\begin{alignedat}{4}
		&\text{$\meg\big|_{\am}$:~}
		&&\LRcouplings{31}&&<
		1.3\cdot 10^{-4}&& \rightarrow 4.9\cdot 10^{-5}\,,\\
		&\text{$\tmg\big|_{\am}$:~}
		&&\LRcouplings{33}&&<
		1.7 &&\rightarrow 0.66\,.
	\end{alignedat}
\end{equation}
They can be understood in two ways.
On the one hand, the perturbativity upper limit together with \am{}
implies individual lower limits on the \am{}-related couplings. Via
Eq.~\eqref{eq:two-top-overall} this translates into the individual upper
limits \eqref{eq:individualtwobody}.
On the other hand, Eq.~\eqref{eq:individual-top-am} implies that $k_{32}$
is bounded. Hence for very large/very small $k_{31}$ the $k$-dependent
factor in Eq.~\eqref{eq:top-two-kk} decreases, again explaining the
upper bounds on individual couplings in 
Figures~\ref{fig:top-meg} and \ref{fig:top-tmg}.

We repeat the previous discussion for the second scenario where only
charm-quark couplings are non-zero. The analysis and conclusions
proceeds analogously to the previous case where top-quark couplings
were non-vanishing.

The four relevant charm-quark couplings for the decay \twodecays{} are
$\LRcouplings{2i}$ and $\LRcouplings{2j}$. The semi-numerical
approximation for the general bound on the combination of these four
couplings reads (all quantities with unit of  mass are to be given in
units of \gev{}; \emph{charm-only scenario}):
\begin{equation}
	\norm{ \Rcoupling{2i} \Lcoupling{2j} } +
	\norm{ \Lcoupling{2i} \Rcoupling{2j} } < 
	\frac{\Gamma_i\br(\twodecays)}{m_i^3}
	\frac{1.2\cdot10^7\ms{4}}{(1-2.4 \ln\ms{})^2} \,.
	\label{eq:two-charm-overall}
\end{equation}
This limit in Eq.~\eqref{eq:two-charm-overall} can be rewritten by
using the ratios $k_{ij}$ between left- and right-handed couplings,
see Eq.~\eqref{eq:ratio-definition}. For the mass $\ms{}=1.8~\tev$ we
obtain (\emph{charm-only scenario})
\begin{equation}
	\begin{alignedat}{2}
		\Lcoupling{21} \Rcoupling{21}\Lcoupling{22} \Rcoupling{22}
		&< 
		1.1\cdot10^2
		~&&\br(\meg) \frac{\ratio{21}\ratio{22}}{k_{21}^2+k_{22}^2}
		\,,\\
		\Lcoupling{23} \Rcoupling{23}\Lcoupling{22} \Rcoupling{22}
		&< 
		1.7\cdot10^5
		~&& \br(\tmg) \frac{\ratio{23}\ratio{22}}{k_{23}^2+k_{22}^2}
		\,.
		\label{eq:charm-two-kk}
	\end{alignedat}
\end{equation}
Combining these upper limits with lower limits on couplings derived
from assuming an explanation of \am{} in
Eq.~\eqref{eq:charm-am-lower-and-upper-limit} yields upper limits on
products of only two couplings relevant for each decay (\emph{charm-only scenario}):
\begin{equation}
	\begin{alignedat}{4}
		&\meg\big|_{\am}:~
		&&\Lcoupling{21} \Rcoupling{21}
		&&< 
		1.2\cdot10^{-10}
		&&\rightarrow
		1.8\cdot10^{-11}
		\,,\\
		&\tmg\big|_{\am}:~
		&&\Lcoupling{23} \Rcoupling{23} 
		&&< 
		2.1\cdot10^{-2}
		&&\rightarrow
		3.2\cdot10^{-3}
		\,.
		\label{eq:charm-am-connected-limits}
	\end{alignedat}
\end{equation}

The corresponding bounds are visualized in the plots of
Figures~\ref{fig:charm-meg} and \ref{fig:charm-tmg}. 
As in the case of the top-quark couplings, the allowed regions
correspond to  essentially hyperbolic shapes as can be understood
from Eq.~\eqref{eq:charm-am-connected-limits}.

Like in the top-coupling case, the figures also show that there are
cutoffs for individual couplings. They arise from the lower limits on
\am{}-related couplings of Eq.~\eqref{charmindividuallimits}.
There are general cutoffs related to
the perturbativity limit combined with requiring a \am{}
explanation. And there are even stronger cutoffs on the left-handed
couplings related to the \kplus{} decay which, via \am{}, implies a
lower limit on \Rcoupling{22}. Numerically, the upper limits on
individual couplings related to the \meg{} and \tmg{} decays read (\emph{charm-only scenario}, individual limits):
\begin{equation}
\begin{aligned}
	\text{$\meg\big|_{\am}$:~}\LRcouplings{21} &<
	1.3\cdot10^{-4} \rightarrow  5.0\cdot 10^{-5}\,,\\
	\Lcoupling{21} &< 4.6\cdot10^{-6} \rightarrow  1.7\cdot 10^{-6}\,,\\
	\text{$\tmg\big|_{\am}$:~}\LRcouplings{23} &< 1.7 \rightarrow 
	0.67\,,
	\\
	\Lcoupling{23} &< 6.0\cdot 10^{-2} \rightarrow 
	2.3 \cdot 10^{-2}\,,
\end{aligned}
\end{equation}
where the second/fourth lines correspond to the constraints from the
\kplus{} decay.


Turning to the third scenario with quark-universal couplings,
the analysis proceeds similar to the previous cases.
We just provide the results, which can also be read off from
Figures~\ref{fig:column-meg} and~\ref{fig:column-tmg}.
The limits on the coupling products are similar to the top-only case
since the top-quark provides the dominant contribution (\emph{columns scenario}):
\begin{equation}
	\begin{alignedat}{4}
		&\meg\big|_{\am}:~
		&&\Lcoupling{1} \Rcoupling{1}
		&&< 
		2.5\cdot10^{-12}
		&&\rightarrow 
		3.6\cdot10^{-13}
		\,,\\
		&\tmg\big|_{\am}:~
		&&\Lcoupling{3} \Rcoupling{3} 
		&&< 
		4.3\cdot 10^{-4}
		&&\rightarrow
		6.7\cdot 10^{-5}
		\,.
	\end{alignedat}
\end{equation}
Similarly, the individual limits from \am{} together with
perturbativity are similar to the case of the top-quark, however the
additional limits from \kplus{} decay are different due to the
combined contributions from up- and charm-quarks (\emph{columns scenario}, individual limits):\footnote{The limits on couplings obtained in this section supersede the ones coming from $\Delta a_e$~\cite{Chen:2022hle, Bigaran:2021kmn} and $\Delta a_\tau$~\cite{DELPHI:2003nah, Burmasov:2022gnl} under the assumption of $\Delta a_\mu$, thus the former are not mentioned in this paper.}

\begin{equation}
	\begin{alignedat}{4}
		&\text{$\meg\big|_{\am}$:~}
		&&\LRcouplings{1} &&<
		1.6\cdot10^{-4} &&\rightarrow  6.1\cdot 10^{-5}
		\,,\\
		&&&\Lcoupling{1} &&<
		1.8\cdot10^{-6} &&\rightarrow  7.0\cdot 10^{-7}
		\,,\\
		&\text{$\tmg\big|_{\am}$:~}&&\LRcouplings{3} &&< 2.1 &&\rightarrow 
		0.83
		\,,\\
		&&&\Lcoupling{3} &&< 2.4 \cdot 10^{-2} &&\rightarrow 
		9.5 \cdot 10^{-3}
		\,.
	\end{alignedat}
\end{equation}

\subsection{Consequences for \teg{}}

Here we consider the decay \teg{}. It is also dipole-induced and
chirality-flip enhanced, but it is not connected to \am{}-related
couplings.
The decay can be analyzed analogously to \meg{} and \tmg{}, and we
present only results for the two generation-specific scenarios. For the
top-only case this leads to the constraint (\emph{top-only scenario})
\begin{equation}
	\begin{alignedat}{4}
		\teg:~
		&\Lcoupling{31} \Rcoupling{31} \Lcoupling{33} \Rcoupling{33} &<
		8.0\cdot10^{-7}
		&&\rightarrow
		2.2\cdot10^{-7}
		\,.
		\label{eq:tegtoplimit}
	\end{alignedat}
\end{equation}
Here, a $k$-dependent factor similar to the ones in
Eq.~\eqref{eq:top-two-kk} has been maximized to obtain the most
conservative bound. We see that the bound involves the same four
couplings as the ones of Eq.~\eqref{eq:top-am-connected-limits}
restricted by \meg{} and \tmg{}, but it is considerably weaker: if
the limits in  Eq.~\eqref{eq:top-am-connected-limits} are met, the
additional bound of Eq.~\eqref{eq:tegtoplimit} is automatically
satisfied by many orders of magnitude.

The analogous result for the case of purely charm-quark couplings reads (\emph{charm-only scenario})
\begin{equation}
	\begin{alignedat}{4}
		\teg:~
		&&\Lcoupling{21} \Rcoupling{21} \Lcoupling{23} \Rcoupling{23} &&<
		2.8\cdot10^{-3}
		&&\rightarrow 
		7.7\cdot10^{-4}
		\,.
	\end{alignedat}
\end{equation}
Again, this limit is many orders of magnitude weaker than the
combination of limits derived from \meg{} and \tmg{} under the
assumption of an explanation of \am{}\ in
Eq.~(\ref{eq:charm-am-connected-limits}).

\section{Phenomenological consequences of three-body decays \meee{} and others}\label{sec:three-body}
\begin{figure}[t!]
	\centering
	\includegraphics[height=\picwidth]{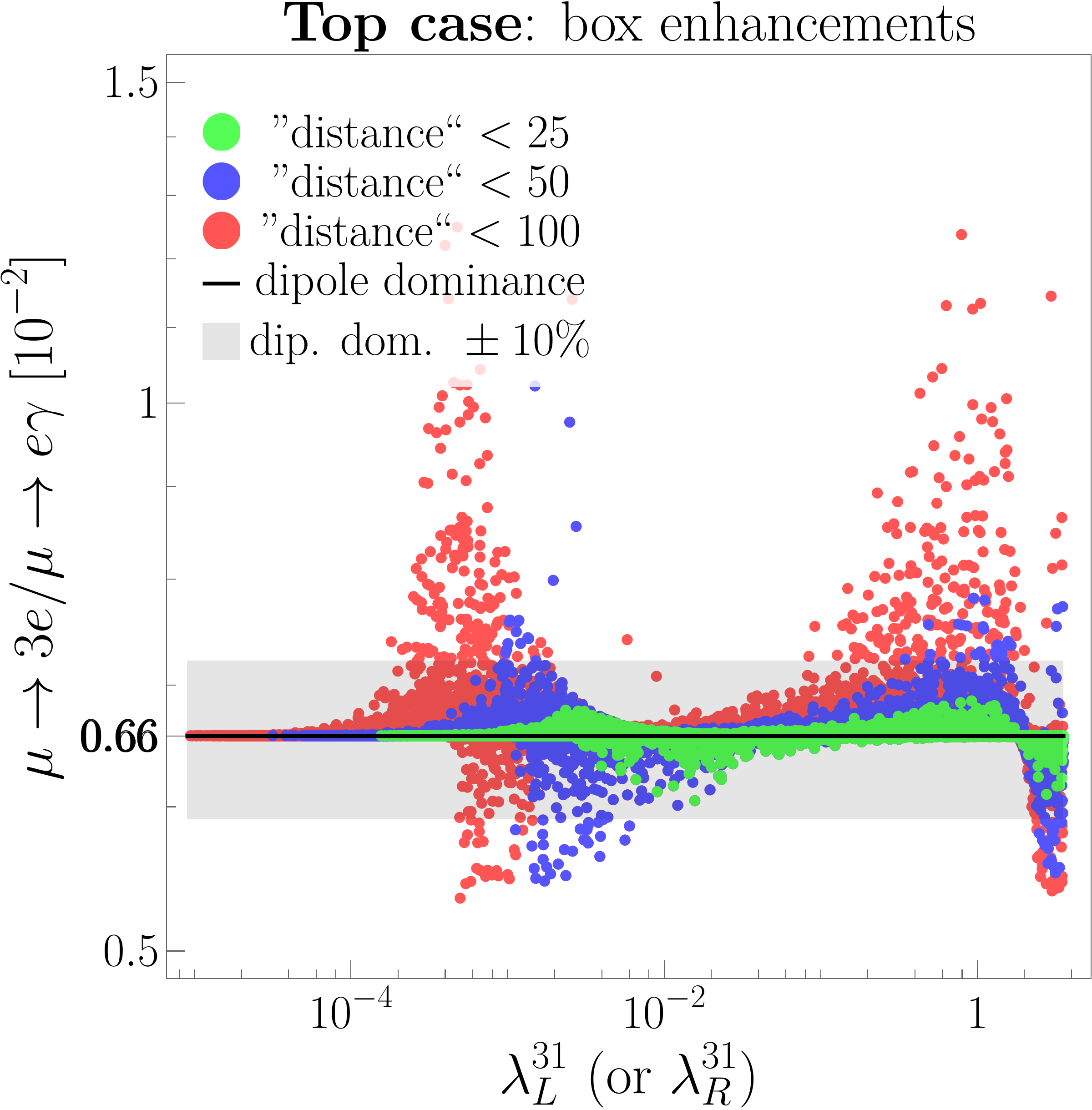}
	\caption{
		\label{fig:ratiomueee}
		Predicted ratios of the branching ratios for
		\meee{} and \meg{}, for a range of
		$\LRcouplings{31}$, resulting from a scan over the relevant
		couplings, for $\ms{}=1.8~\tev$. The gray band shows the value of
		equation~\eqref{eq:meeedipoledominance} for dipole dominance (with a
		$\pm10\%$ corridor), and the color code of points corresponds to the
		spread between the four relevant couplings $\LRcouplings{31}$ and
		$\LRcouplings{32}$ in case of scenario 1. This ``distance'' is computed
		by taking the four relevant couplings, calculating their geometric
		mean, and then determining the maximum difference to the mean,
		divided by the mean.
	}
\end{figure}

The phenomenological discussion of three-body decays \threedecays{}, particularly of
\meee{}, can be kept brief. Even though these processes are influenced
by a variety of vertex and box form factors, they are strongly
dominated by the dipole form factors $A_2^{L,R}$ in those parts of
parameter space which gives rise to conservative bounds aimed for in
the present study. For this purpose the
three-body decays are strongly correlated to the simpler two-body
decays \twodecays{}.

The dipole dominance is illustrated in Figure~\ref{fig:ratiomueee},
which shows the ratio of the two predicted branching ratios for
the two most interesting processes \meee{} and \meg{}, for a range of
$\LRcouplings{}$. The color code of the points corresponds to the
spread between the four relevant couplings $\LRcouplings{31}$ and
$\LRcouplings{32}$ in case of scenario 1.
If the spread is moderate (green/blue points),
we have an essentially fixed ratio between the
branching ratios for \meee{} and \meg{}, which is approximately
\begin{equation}
	\frac{\br(\meee)}{\br(\meg)}  = 6.6\cdot 10^{-3}\,.
	\label{eq:meeedipoledominance}
\end{equation}
The dipole dominance in this parameter region has two reasons.

First, the photonic form factor $A_1^{L,R}$ behaves similarly to the
non-chirally enhanced terms in the dipole form factor
$A_2^{L,R}$. Since we are working in a coupling regime with strong
chirality-flip enhancements, see
Eq.~\eqref{eq:chiralenhancementcriterion}, $A_1^{L,R}$ provides only a
negligible correction.

Second, the box diagrams (giving rise to contributions to vector,
scalar and tensor form factors) are in principle of general interest since they
depend on four powers of $\LRcouplings{}$. However, if the spread
between the couplings is moderate this cannot lead to
enhancements, resulting in the correlation
(\ref{eq:meeedipoledominance}).

However, if a large spread is allowed (red points in
Figure~\ref{fig:ratiomueee}), the behavior is more
complicated and either enhancement or destructive interference is
possible.

On the one hand, the derivation of conservative bounds in the style of
the figures of Sec.~\ref{sec:two-body} depends on the parameter
points without box enhancements. Hence, 
given the available experimental limits of
Table~\ref{tab:observables-experiments}, the
three-body decays do not provide additional constraints on top of the
ones obtained from two-body decays analyzed in the previous
section. This remains true even for  the next phases of the
experiments listed in Table~\ref{tab:observables-experiments}.

On the other hand, the enhanced red points show that future
\meee{} measurements are promising since enhanced rates are possible
in this leptoquark model. Finally, the planned 
Mu3e-II~\cite{Mu3e:2020gyw} experiment for \meee{}, which we otherwise
do not consider in the
present paper has significant potential for discovery and for
improvements of bounds even on the dipole form factors beyond the
limits presented in Sec.~\ref{sec:two-body}.

\section{Phenomenological consequences of $\mu\to e$ conversion}\label{sec:conversion}
\begin{figure*}[t!]
	\centering
	\begin{subfigure}[b]{\picwidth}
		\includegraphics[height=\textwidth]{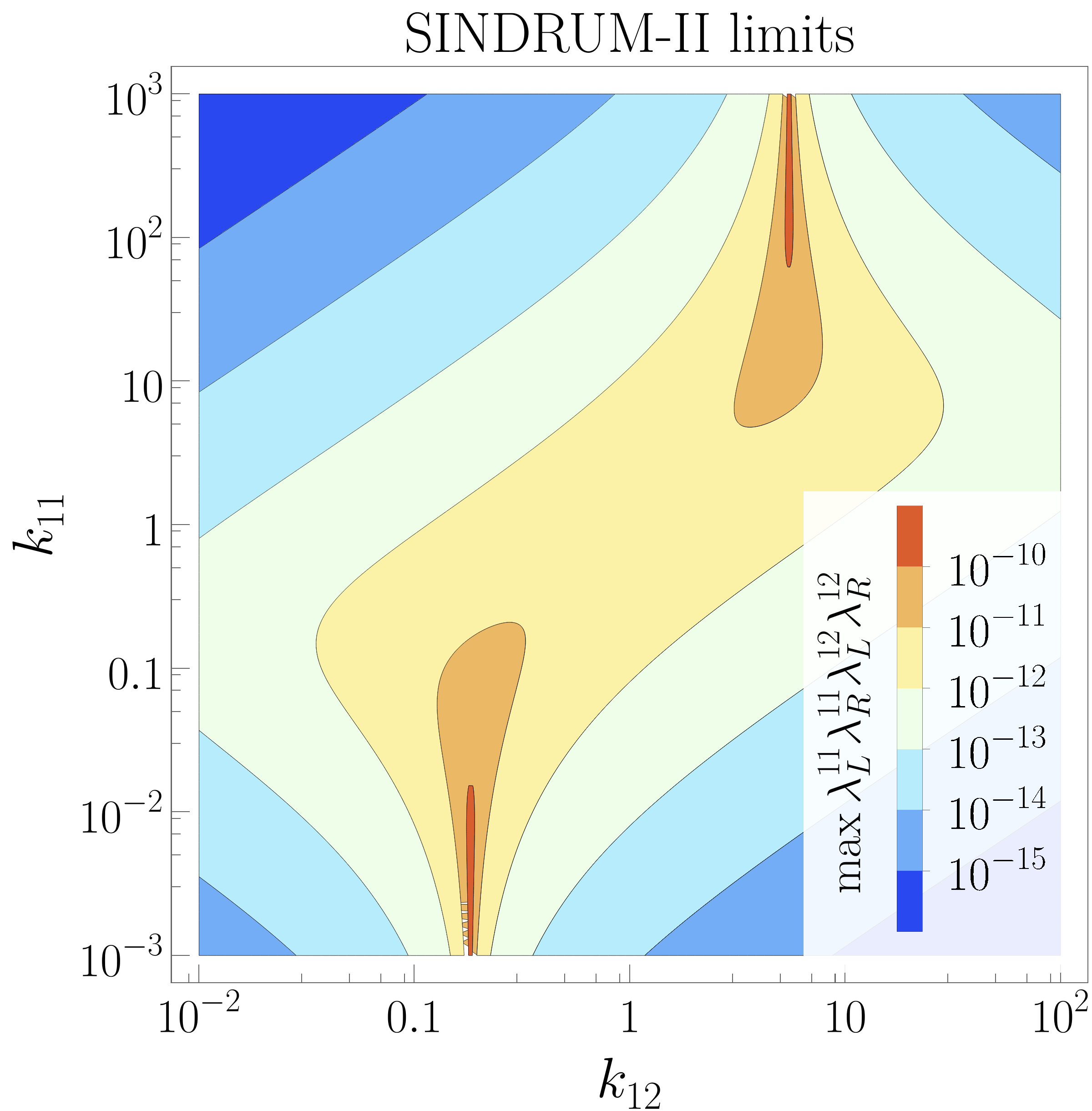}
		\caption{}
		\label{fig:mec-kk}
	\end{subfigure}
	\begin{subfigure}[b]{\picwidth}
		\includegraphics[height=\textwidth]{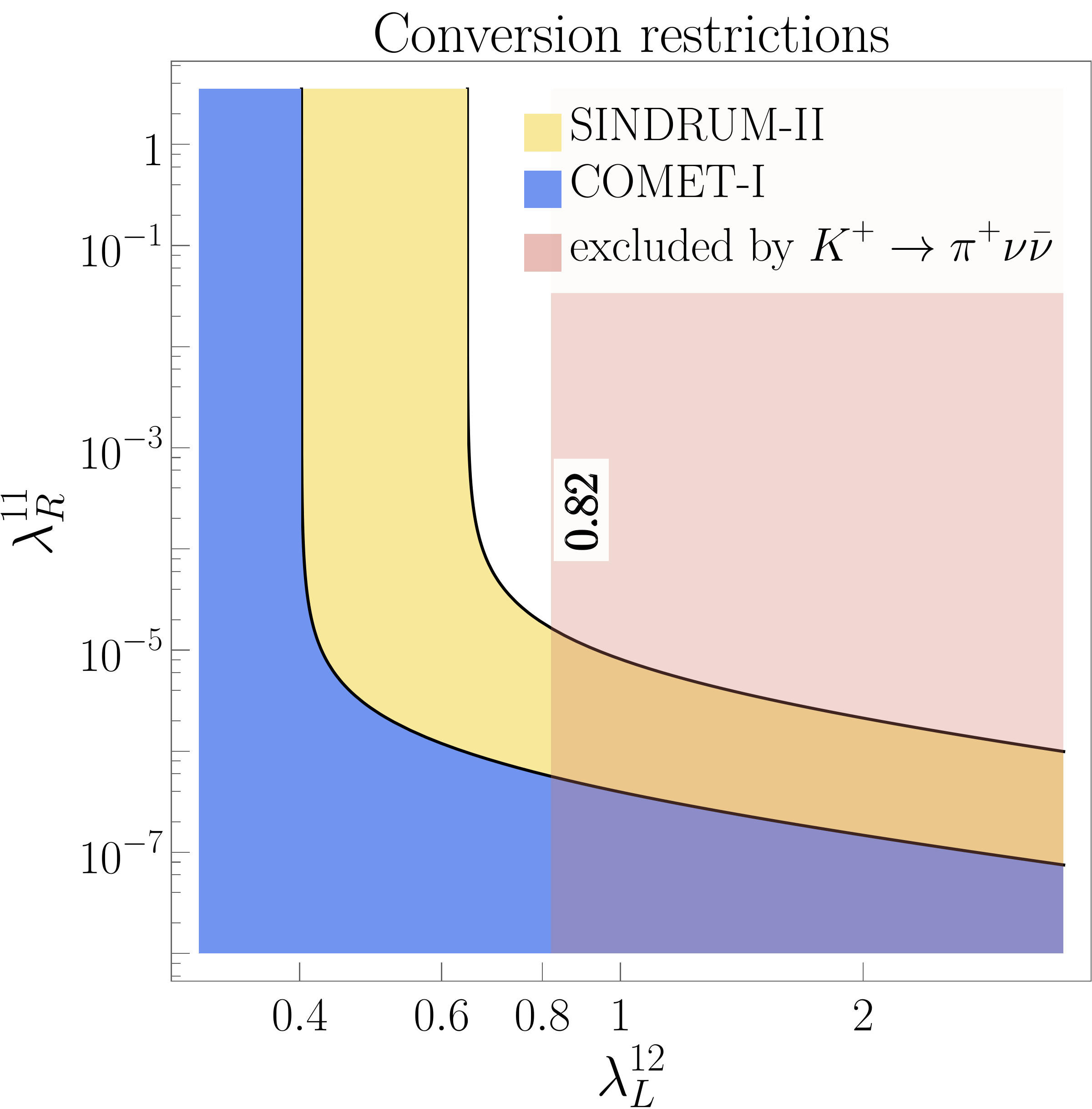}
		\caption{}
		\label{fig:free-conversion}
	\end{subfigure}
	\caption{
		Limits for \mecold{} (SINDRUM-II) and \mec{} (COMET-I)
		for $\ms{}=1.8~\tev$ and different coupling patterns. For
		Figure~\ref{fig:mec-kk} see equation \eqref{eq:mec-kk}, and
		for Figure~\eqref{fig:free-conversion} 
		see Eq.~\ref{eq:conversion-restriction}. The red
		shaded area is fully excluded by the \kplus{} decay.
	}
\end{figure*}

To discuss the phenomenological impact of $\mu-e$ conversion process we
rewrite the leptoquark contribution in a way similar to the previous
observables, as a product of the four relevant couplings and a
$k_{ij}$-dependent factor, as
\begin{equation}
	\br(\mu-e) = \frac{\alpha_{s}^2}{4\ms{4}\omega_{\mathrm{capt}}} 
	\Lcoupling{11}\Rcoupling{11}\Lcoupling{12}\Rcoupling{12}
	~k_\alpha\,.
	\label{eq:BRmec-kk}
\end{equation}
This highlights that the four relevant couplings are
$\LRcouplings{11}$ and $\LRcouplings{12}$. The dimensionless
$k_{ij}$-dependent factor is denoted as $k_\alpha$; it is more
involved than for previous cases and this time also depends on the
form factors $\alpha_{s,v}$,
\begin{equation} 
	k_\alpha=
	\frac{\big( \ratio{12} - \frac{\alpha_{v}}{\alpha_{s}} \big)^2}{\ratio{11}\ratio{12}}
	+
	\ratio{11}\ratio{12}\left(\frac{1}{ \ratio{12}} - \frac{\alpha_{v}}{\alpha_{s}} \right)^2
	\,.
	\label{eq:mec-kalpha}
\end{equation}

Similarly to the previous observables we can obtain a limit on the product of
the four relevant couplings, now depending on the
factor $k_\alpha$. Given the present experimental bound from the SINDRUM-II
experiment (or the expected bounds from COMET-I), this limit reads. Note, that form factors $\alpha_{s,v}$ in $k_\alpha$ should be taken appropriately to the nucleus from Eq.~\eqref{eq:mec-formfactors}.

\begin{equation}
	\Lcoupling{11}\Rcoupling{11}\Lcoupling{12}\Rcoupling{12}
	<
	\frac{8.4\cdot 10^{-12}}{k_\alpha^{Au}}
	\to
	\frac{5.6\cdot 10^{-14}}{k_\alpha^{Al}}
	\,.
	\label{eq:mec-kk}
\end{equation}
Figure~\ref{fig:mec-kk} displays this limit for the case of the
present bound from the SINDRUM-II experiment. In the figure,
the color code corresponds to the upper limit on the coupling product,
on the axes the two ratio variables \ratio{11} and \ratio{12} are
varied. The shape of the figure can be explained as follows.

Within the $k$-dependent
factor there can be cancellations: if either
$\ratio{12}=\alpha_{v}/\alpha_{s}$ or 
$\ratio{12}=\alpha_{s}/\alpha_{v}$, the prefactor of the first (or second)
term in Eq.~\eqref{eq:mec-kalpha} vanishes. If simultaneously \ratio{11}
becomes very small (or large), the entire factor $k_\alpha$ is very
small, and conversely very large coupling products are allowed. This
explains the two horizontal strips in the figure where the limit
becomes significantly weaker.

Given this complicated behavior, it is instructive to record the limit
in some special cases with different degree of possible cancellations. 
First, in the special point where $\ratio{11}=\ratio{12}=1$, i.e.~where the
left- and right-handed couplings happen to be equal, the limits become

\begin{equation}
	\begin{alignedat}{1}
		\mecold|_{\Lcoupling{}=\Rcoupling{}}:&\quad
		\LRcouplings{11}\LRcouplings{12}<2.5\cdot10^{-6}
		\,,\\
		\mec|_{\Lcoupling{}=\Rcoupling{}}:&\quad
		\LRcouplings{11}\LRcouplings{12}<1.9\cdot10^{-7}
		\,.
	\end{alignedat}
\end{equation}
Second, we consider  the region where the left- and right-handed
couplings may
differ by up to a factor 10, $\ratio{11},\ratio{12}\in[0.1,10]$. In this
region one of the terms within $k_\alpha$ can vanish, and overall
$k_\alpha$ turns out to vary in the interval
$k_\alpha=0.48\ldots96$ ($0.73\ldots98$ for COMET-I). 
A limit on the coupling product which is valid in all of the
region for $\ms{}=1.8~\tev$ reads:
\begin{equation}
	\Lcoupling{11}\Rcoupling{11}\Lcoupling{12}\Rcoupling{12}<
	6.5\cdot10^{-12}
	\to
	3.7\cdot10^{-14}
	\,.
\end{equation}

The previous observables have allowed (in conjunction with \am{}) to
obtain bounds on individual couplings which are complementary to the
bounds on coupling products. This is more difficult in the case
of \mec{} or \mecold{}. A major reason is the possibility of
cancellations due to the two terms involving $\alpha_{s}$ and
$\alpha_{v}$. 
It is, however, possible to obtain rather strict limits on the
correlation of a subset of two
couplings. This is illustrated in Figure~\ref{fig:free-conversion},
which shows the allowed regions in the plane of the two couplings
$\Lcoupling{12}$--$\Rcoupling{11}$. The remaining two parameters have
been scanned over. (A similar plot could be shown  in
the $\Rcoupling{12}$--$\Lcoupling{11}$ plane.)

To explain the shape of the plot it is useful to 
discuss Eq.~\eqref{eq:mec-overall} (or Eq.~\eqref{eq:mec-kalpha})
distinguishing two cases for the couplings: either we have
$\Lcoupling{12} > \tfrac{\alpha_v}{\alpha_s}\sqrt{4\pi} 
$ or we have $\Lcoupling{12} \le \tfrac{\alpha_v}{\alpha_s}\sqrt{4\pi}
$.
In the first case, no matter what the value of $\Rcoupling{12}$ is,
the prefactor of $\Rcoupling{11}$ in the branching ratio is not zero;
hence we get an upper limit on $\Rcoupling{11}$. In the second case,
there is a certain value of $\Rcoupling{12}$ (within the perturbative
regime) which nullifies the prefactor of $\Rcoupling{11}$; hence that
latter coupling can be arbitrarily large. This behaviour explains the
shape of the allowed regions in the
plot.  The upper limit on $\Rcoupling{11}$ can also be described by the formula
\begin{equation}
\begin{gathered}
	\Rcoupling{11}  < \frac{2\ms{2}\sqrt{\omega_{\mathrm{capt}}\br(\mu-e)}}
{\alpha_s \Lcoupling{12}-\sqrt{4\pi}\alpha_v}
~\text{if}~
\Lcoupling{12} > \tfrac{\alpha_v}{\alpha_s}\sqrt{4\pi}
\,,
\label{eq:conversion-restriction}
\end{gathered}
\end{equation}
which is valid with $L\leftrightarrow R$ replacement and again explains the shape of the plot.

\section{Conclusions}\label{sec:conclusions}

\begin{table*}[t!]
	\newcommand{\textsize}{\scriptstyle}
	\centering
	\newcommand{\less}{\,<\,}
	\newcommand{\ra}{\,\rightarrow\,}
	\setlength{\tabcolsep}{5pt}
	\begin{tabular}{| c | c | c | c | c |}
		\hline
		$q\backslash \ell$ & $e$ & $\mu$ & $\tau$ & valid
		\\\hline
		\multirow{3}{*}{$u$}
		&\multicolumn{2}{c|}{%
			$\textsize\Lcoupling{11}\Rcoupling{11}\Lcoupling{12}\Rcoupling{12}\less
			6.5\cdot10^{-12} \ra 3.7\cdot10^{-14}$}
		&
		\multirow{3}{*}{---}
		&\multirow{3}{*}{any sc.}
		\\\cline{2-3}
		&
		$\textsize\Lcoupling{11}(\Rcoupling{12}-0.65) \less 2.9\cdot 10^{-6} \ra$
		&
		\multirow{2}{*}{$\textsize\Lcoupling{12} \less  0.82$}
		&&
		\\
		&
		$\textsize\Lcoupling{11}(\Rcoupling{12}-0.40) \less 2.4\cdot 10^{-7}$
		&&&
		\\\hline
		\multirow{3}{*}{$c$} &
		$\textsize\Lcoupling{21}\Rcoupling{21} \less 1.2\cdot 10^{-10} \ra 1.8\cdot 10^{-11}$
		&
		$\textsize0.18 \less \Lcoupling{22}\Rcoupling{22} \less 0.56$
		&
		$\textsize\Lcoupling{23}\Rcoupling{23} \less 2.1\cdot 10^{-2} \ra 3.2\cdot 10^{-3}$
		& \multirow{3}{*}{sc. 2}\\&
		$\textsize\LRcouplings{21} \less 1.3\cdot10^{-4} \ra 5.0\cdot 10^{-5}$
		&
		$\textsize5.1\cdot10^{-2} \less \LRcouplings{22} \less \sqrt{4\pi}$
		&
		$\textsize\LRcouplings{23} \less 1.7 \ra 0.67$
		&\\&
		$\textsize\Lcoupling{21} \less 4.6\cdot 10^{-6} \ra 1.7\cdot10^{-6}$
		&
		$\textsize\Lcoupling{22} \less 0.13\,,~1.5 \less \Rcoupling{22}$
		&
		$\textsize\Lcoupling{23} \less 6.0\cdot10^{-2} \ra 2.3\cdot10^{-2}$
		&
		\\\hline
		\multirow{2}{*}{$t$} &
		$\textsize\Lcoupling{31}\Rcoupling{31} \less 2.1\cdot10^{-12} \ra 2.9\cdot 10^{-13}$
		&
		$\textsize3.1\cdot10^{-3} \less \Lcoupling{32}\Rcoupling{32} \less 9.3\cdot10^{-3}$
		&
		$\textsize
		\Lcoupling{33}\Rcoupling{33} \less 3.5\cdot10^{-4} \ra 5.4\cdot 10^{-5}$
		& \multirow{2}{*}{sc. 1}\\&
		$\textsize\LRcouplings{31} \less 1.3\cdot 10^{-4} \ra 4.9\cdot 10^{-5}$
		&
		$\textsize8.7\cdot10^{-4} \less \LRcouplings{22} \less \sqrt{4\pi}$
		&
		$\textsize\LRcouplings{33} \less 1.7 \ra 0.66$
		&\\
		\hline
	\end{tabular}
	\caption{
		Summary of restrictions on all entries of the $S_1$
		leptoquark coupling matrices 
		$\LRcouplings{}$ for $\ms{} = 1.8~\tev{}$. The
		restrictions in the second and third rows are   valid under the condition that
		$\am{}$ of equation~(\ref{eq:Damu}) is explained, and
		they apply to
		various scenarios of Sec.~\ref{sec:strategy} as
		indicated in the rightmost columns.
		For the derivation and the range of validity of the
		constraints on individual couplings we refer to the
		appropriate sections and text.
	} 
	\label{tab:all-restrictions}	
\end{table*}

In the present paper, we have analyzed the impact of combining $\am{}$
with CLFV limits on the parameter space of the $S_1$ leptoquark
model. This well-motivated model involves two $3\times3$ coupling
matrices $\LRcouplings{q\ell}$ whose entries are strongly constrained
by the combination of low-energy lepton observables. Here we briefly
summarize and comment on the most important results.

The summary is also displayed in Table~\ref{tab:all-restrictions} in a
matrix form, such that the $q\ell$-entry of
Table~\ref{tab:all-restrictions} collects constraints on  the entries
$\LRcouplings{q\ell}$.

Generally $\am{}$ from equation~(\ref{eq:Damu}) implies upper and
lower limits on the left-right products of couplings to muons, and
CLFV constraints then lead to upper limits on left-right products of
couplings to the electron and $\tau$ lepton. In addition,
perturbativity and the \kplus{} decay imply upper limits on individual
couplings; these (together with limits on products) produce also lower
limits on other individual couplings.

Specifically the third row of  Table~\ref{tab:all-restrictions}
assumes the ``top-only'' scenario (see section \ref{sec:strategy})
where $\am{}$ is explained via top-quark couplings 
only. In this case the (geometric average of left- and right-handed)
couplings to electrons must be more 
than 4 orders of magnitude smaller than the corresponding couplings to
muons. Also, the couplings to $\tau$ leptons must be smaller than the
couplings to muons. In the absence of cancellations within the theory
predictions, this conclusion remains unchanged even in the more general case where couplings to the charm- and up-quarks are also
allowed to be nonzero (but small so as to not significantly modify the
contributions to $\am{}$).

Similarly, the second row of Table~\ref{tab:all-restrictions}
assumes the ``charm-only'' scenario and presents bounds on couplings
of leptons to the charm-quark. In order to accommodate the current
$\am{}$ value, the couplings to the muon must be ${\cal O}(1)$. In
addition, 
the \kplus{} decay implies limits on the ratio of left- and
right-handed couplings, valid in a wide range of
parameter space (see section \ref{sec:am} for details). These are also
reflected in the asymmetries visible in
Figures~\ref{fig:charm-meg}~and~\ref{fig:charm-tmg} for the \meg{} and
\tmg{} decays.
Again, there
must be a strong hierarchy between charm-couplings to the muon and to
the electron.

Finally, the first row of  Table~\ref{tab:all-restrictions}
is valid irrespective of the scenario. It is derived from $\mu\to e$
conversion constraints and from the \kplus{} decay. As a result of these constraints, the
(geometric average of the) couplings of electrons and muons to the
up-quark must be significantly smaller than the couplings to the
charm- or top-quarks if the $\am{}$ deviation is accommodated.
In addition, more detailed limits on $\Lcoupling{12}$ and on products
of two couplings can be given as shown in the Table~\ref{tab:all-restrictions} and as explained
in section \ref{sec:conversion}.

The table also collects the possible improvements of limits from the
next phases of CLFV experiments collected in Table~\ref{tab:observables-experiments}. If no signal is found, they
will significantly sharpen the upper limits on couplings to electrons
and $\tau$ leptons and will increase the need for highly hierarchical
and non-universal entries in the coupling matrices
$\LRcouplings{q\ell}$. In general, the results exemplify the
implications of $\am{}$ and CLFV constraints on the flavor structure of
new physics models with enhanced chirality flips. Concrete models of
flavor need to be compatible with such results. This is of particular
interest in the considered case of leptoquarks, where an obvious
and unambiguous notion of minimal flavor violation is not available
\cite{Davidson:2006bd,Davidson:2010uu}.

\begin{acknowledgments}
U.Kh. was supported by the Deutscher Akademischer Austauschdienst (DAAD) under Research Grants \textemdash{} Doctoral Programmes in Germany, 2019/20 (57440921)
and by the Deutsche Forschungsgemeinschaft (DFG) under grant number STO 876/7-1.
\end{acknowledgments}
	
\appendix
		
\section{Constraints from flavor-conserving meson decays}
In this appendix, we discuss the impact of two lepton flavor-conserving
decays on the leptoquark parameter space. Both decays have been used
in Ref.~\cite{Kowalska:2018ulj} to constrain the case of charmphilic
explanations of $\am{}$. For earlier, original calculations and
analyses of further meson decays within leptoquark models see
Refs.\ \cite{deBoer:2015boa,Fajfer:2015mia,Cai:2017wry}. Here we
generalize the results of Ref.~\cite{Kowalska:2018ulj} to the case of
general coupling structures.

{
	\subsection{Decay \kplus{}}\label{app:kplus}
	\begin{figure}[t!]
		\centering
		\includegraphics[width=\picwidth]{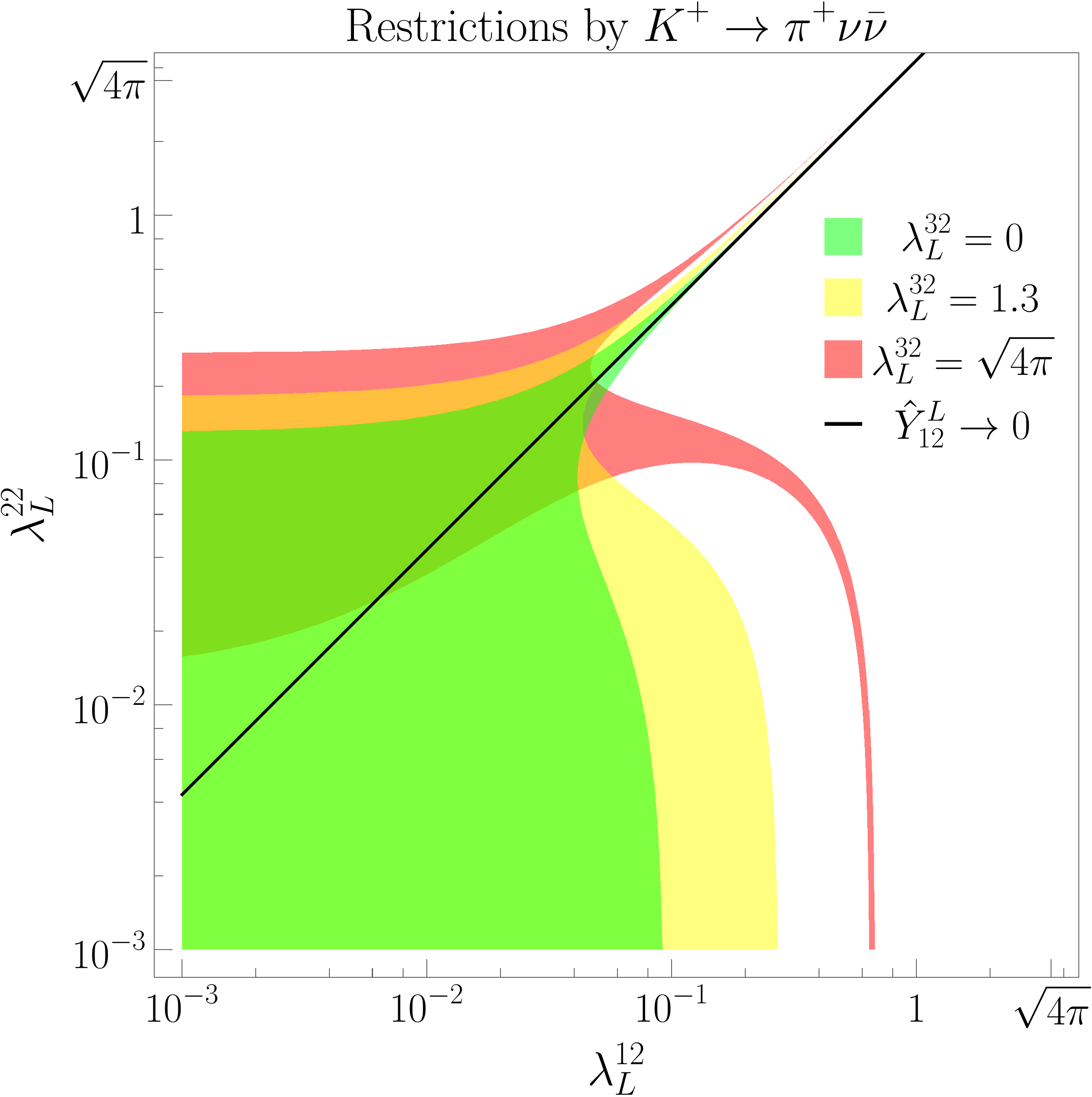}
		\caption{
			Constraints from the \kplus{} decay in the
			$\Lcoupling{12}$--$\Lcoupling{22}$ plane, for different values of
			$\Lcoupling{32}$. The colored regions are allowed for the indicated
			values of $\Lcoupling{32}$.
		}
		\label{fig:kaon-solo}
	\end{figure}
	\newcommand{\coeff}{k}
	From Ref.\ \cite{Kowalska:2018ulj} one obtains the following leptoquark contribution to the branching ratio
	\begin{equation}
		\br(\kplus)_{\text{LQ}}
		=
		\coeff_1 C^{K1\ell}_{VLL}(C^{K1\ell}_{VLL} + \coeff_2)
		\,,
	\end{equation}
	where the following abbreviations are used
	\begin{equation}
		\begin{alignedat}{4}
			C^{K1\ell}_{VLL} &= \frac{1}{2\ms{2}} \hat Y^L_{12} \hat Y^{L*}_{22}
			\,,\qquad
			\hat Y^L_{ql} = V_{\text{CKM}}^{iq}\Lcoupling{il}
			\,,\\
			\coeff_1 &= \frac{\kappa_+}{3C_F^2 \lambda^{10}} \approx 1.83595\cdot10^9~\gev^4
			\,,\\\
			\coeff_2 &= 2C_F\big|\operatorname{Re}[\lambda_t] X(m_t^2/m_W^2)
			+\lambda_c X_{NL}^e\big|
			\\
			&\approx
			2.65751\cdot 10^{-10}~\gev^{-2}
		\end{alignedat}
	\end{equation}
	with the numerical values of intermediate parameters as in
	Ref.~\cite{Kowalska:2018ulj}; after subtracting the SM branching ratio
	($\br(\kplus)_{\text{SM}}\approx 9\cdot 10^{-11}$) from the
	experimental limit obtained by the E949 Collaboration~\cite{E949:2008btt} one obtains the following 2$\sigma$ bounds:
	\begin{equation}
		-1.27  \cdot 10^{-10} < \br(\kplus)_{\text{LQ}} < 3.13 \cdot 10^{-10}\,.
	\end{equation}
	The \kplus{} decay thus constrains a combination of the three
	left-handed parameters $\Lcoupling{i2}$ ($i=1,2,3$). The numerical
	result is shown in Figure \ref{fig:kaon-solo} in the plane of
	$\Lcoupling{12}$--$\Lcoupling{22}$ ($\Lcoupling{32}$ is less important
	since it appears only multiplied with small CKM matrix elements).
	
	The green area corresponds to the allowed region for the special case
	$\Lcoupling{32}=0$. There is a  thin allowed strip which is always
	allowed as $\hat Y^L_{12}$ vanishes due to different signs of CKM matrix
	entries. This strip is cut off only by the  perturbativity limit.
	
	If $\Lcoupling{32}$ is allowed to be nonzero, the allowed region in
	the $\Lcoupling{12}$--$\Lcoupling{22}$ plane can increase. The yellow
	area corresponds to the choice  $\Lcoupling{32}=1.3$. Here the allowed
	region has a similar shape as the green region but extends to larger
	coupling values. If $\Lcoupling{32}$ is increased further, the shape
	of the allowed region changes. The reason is that specific values of
	the up- and charm-quark-couplings are required to cancel the large
	top-coupling contributions. The red region illustrates this for the
	value of $\Lcoupling{32}$ at the perturbativity limit. This region
	also illustrates the absolute achievable upper limit
	\begin{align}
		\Lcoupling{12}< 0.82\,,
	\end{align}
	which is used in Figure~\ref{fig:free-conversion}.
}

{
	\subsection{Decay \dzero{}}\label{appendix:dzero}
	\newcommand{\coeff}{d}
	
	The current experimental bound is the following~\cite{LHCb:2013jyo}:
	
	\begin{equation}
		\br(\dzero) < 7.6\cdot 10^{-9}~(95\%~\text{CL})
		\,.
	\end{equation}
	
	The expression for the branching ratio has the form:
	
	\begin{equation}
		\begin{alignedat}{5}
			\br(\dzero) =\,& \frac{\coeff_1}{\ms{4}}\Big[
			\big(\Lcoupling{12}\Rcoupling{22}-\Rcoupling{12}\Lcoupling{22}\big)^2
			\\&+
			\big(
			\Lcoupling{12}\Rcoupling{22}+\Rcoupling{12}\Lcoupling{22}
			\\&+\coeff_2(\Lcoupling{12}\Lcoupling{22}+\Rcoupling{12}\Rcoupling{22})
			\big)^2
			\Big]
		\end{alignedat}
	\end{equation}
	with the following abbreviations and numerical values from~\cite{Kowalska:2018ulj}:
	\begin{equation}
		\begin{alignedat}{5}
			&\coeff_1 = \tau_D \frac{f_D^2}{256\pi} \frac{m_D^5}{m_c^2}
			\approx
			(17.3~\gev)^4
			\,,\\& 
			\coeff_2 = \frac{m_\mu m_c}{m_D} 
			\approx 0.0391
			\,.
		\end{alignedat}
	\end{equation}
	This decay leads to relevant constraints for the down-type
	coupling basis considered in 
	Ref.~\cite{Kowalska:2018ulj}. For our purposes, we employ the up-type
	basis and several scenarios as described in section
	\ref{sec:strategy}. We have checked that for all our scenarios this
	decay does not lead to additional bounds on parameter space beyond the
	bounds presented in the main text of the paper.

\bibliography{bibliography}
		
\end{document}